\documentclass[12pt]{article}
\usepackage[margin=1in]{geometry}
\usepackage{graphicx}
\usepackage[utf8]{inputenc}
\usepackage[T1]{fontenc}
\DeclareUnicodeCharacter{02BB}{\textquoteleft}
\usepackage{amsmath}
\usepackage{authblk}

\usepackage{hyperref}
\usepackage[font=small,labelfont=bf]{caption}
\usepackage[authordate,backend=biber,dashed=false]{biblatex-chicago}
\title{Coloring Black Holes: Epistemic and Aesthetic Choices in Astronomical Imaging}
\author[1]{Rodrigo Ochigame\thanks{\texttt{rodrigo@ochigame.org}}}
\author[2,3,4]{Emilie Skulberg\thanks{\texttt{emilie.skulberg@nbi.ku.dk}}}
\author[3,4]{Jeroen van~Dongen\thanks{\texttt{j.a.e.f.vandongen@uva.nl}}}
\affil[1]{Institute of Cultural Anthropology and Development Sociology, Leiden University}
\affil[2]{Niels Bohr Archive, University of Copenhagen}
\affil[3]{Institute for Theoretical Physics Amsterdam, University of Amsterdam}
\affil[4]{Vossius Center for the History of Humanities and Sciences, University of Amsterdam}
\date{}

\begin{document}

\maketitle

\begin{abstract}
\noindent 
In 2019, the first image of a black hole's shadow based on observation was released by the Event Horizon Telescope Collaboration (EHT). This paper shows that despite the EHT's emphasis on a single image as its final result, there were countless plausible ways of rendering the data, among which researchers could not easily choose. To obtain a single image from the extremely noisy and sparse data, it was necessary to select one of multiple plausible approaches, or to average the results from different approaches, at each stage of data processing. We examine the epistemic and aesthetic choices involved at various stages, and explore what the images would have looked like if the EHT had made different choices. We suggest that the most valuable evidence produced by the EHT comes not from the single image it ultimately advertised as its central result, but from the demonstration of the limited variability that emerged from the specific choices made.
\end{abstract}

\section{Introduction} \label{introduction}

\begin{figure}[t]
    \centering
    \includegraphics[width=\textwidth]{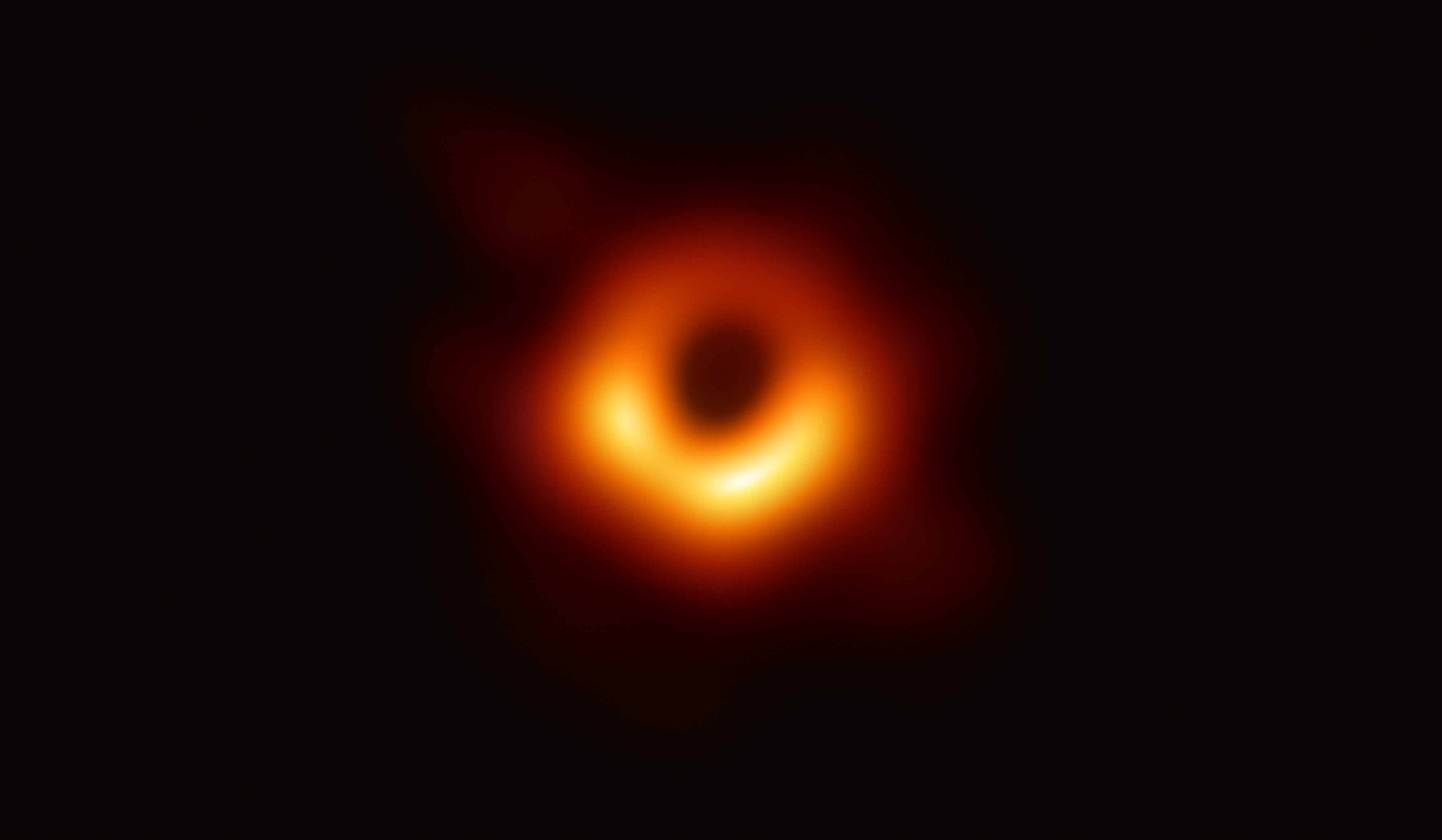}
    \caption{The first image of the shadow of a black hole (M87*) on the basis of observation. Credit: Event Horizon Telescope Collaboration.}
    \label{fig:m87-public}
\end{figure}

On April 10, 2019, the Event Horizon Telescope Collaboration (EHT) released the first image of a black hole's shadow based on observation. At the US press conference, EHT founding director Sheperd Doeleman introduced Figure~\ref{fig:m87-public} by expressing that he and his collaboration were ``delighted to be able to report to you today that we have seen what we thought was unseeable. We have seen---and taken a picture of---a black hole. Here it is.''\footnote{\textcite[timestamp: 7:30]{event_horizon_telescope_collaboration_and_national_science_foundation_nsf_2019}.} ``Taken a picture'' might imply a straightforward process. Yet the journey from data collection during the 2017 observing campaign to the publication of results lasted two years. The making of the image shown in Figure~\ref{fig:m87-public} entailed significant labor and involved choices about data treatment, imaging, and image presentation. This paper unpacks this process and analyzes the steps in the making of Figure~\ref{fig:m87-public} from a historical-epistemological perspective. 

We show that despite the EHT's emphasis on a single image as its final result, there were countless plausible ways of rendering the data, among which researchers could not easily choose. To obtain a single image from the extremely noisy and sparse data, it was necessary to select one of multiple plausible approaches, or to average the results from different approaches, at each stage of data processing. Yet understanding the variability of the results is crucial to fully appreciating the complexity and ambivalent nature of the scientific evidence attained. We suggest that the most valuable evidence produced by the EHT comes not from the single image it ultimately advertised as its central result, but from the demonstration of the limited variability that emerged from the specific choices made.

Black holes are regions where spacetime is warped so extremely that, according to Einstein's general theory of relativity, electromagnetic radiation such as light and radio signals cannot escape from them. They also redirect the radiation emitted by the charged plasma surrounding them, as well as ambient light that originated at other locations in the universe. This led to the prediction of the possible observation of a particular ``shadow'' if a black hole were viewed from Earth.\footnote{\textcite{falcke_viewing_2000}.} But since a black hole itself is `black' and these light signals are notoriously weak, it was also immediately clear that making such an image would be quite a challenge, even if one were to try to `image' some of the closest supermassive black holes, such as the one at the center of the nearby Messier 87 (M87) galaxy, known as M87*, or the one at the center of our own Milky Way, Sagittarius A*.

When the first observational image of the shadow of a black hole (M87*) was presented, the EHT argued that it provided the most convincing evidence to date for the existence of supermassive black holes at the centers of galaxies. The way EHT researchers argued for this depended on their audience. When targeting non-specialists, the EHT used a single image that was often identified as a `photograph,' `photo,' or `picture' to present its results as evidence for existence. In the peer-reviewed papers tied to the 2019 release of the M87* image, the EHT instead described in detail an extensive process of developing images and improving imaging algorithms, and some of the different images that had been part of that process.\footnote{\textcite{event_horizon_telescope_collaboration_first_2019-1}; \textcite{event_horizon_telescope_collaboration_first_2019-2}; \textcite{event_horizon_telescope_collaboration_first_2019-3}; \textcite{event_horizon_telescope_collaboration_first_2019-4}; \textcite{event_horizon_telescope_collaboration_first_2019-5}; \textcite{event_horizon_telescope_collaboration_first_2019-6}.}  Here, the care and precautions taken by the EHT were used to argue that its images provided the best evidence for existence. Instead of implying a quick and straightforward process, the authors emphasized the use of different imaging algorithms, a two-stage imaging process, and how they had carefully double-checked their own methods.

The EHT presented its result as the ``first image of a black hole,'' in the sense of an image from empirical \emph{observation}. For decades, scientists and artists had produced various images of black holes through handmade drawings, computer visualizations, or hybrids thereof; many of these were based on numerical \emph{simulations} of black hole physics.\footnote{For early images in the physics literature, see \textcite{godfrey_machs_1970}; \textcite{bardeen_timelike_1973}; \textcite{cunningham_optical_1973}. On black holes in art, see \textcite{gamwell_conjuring_2025}.} Such simulations predicted that the light from the accretion disk around a black hole would not show up symmetrically, but with one side brighter than the other.\footnote{For the first example of this, see \textcite{luminet_image_1979}. For a historical account of Luminet's work, see \textcite{skulberg_black_2023}. For analyses drawing on semiotics, see \textcite{dondero_images_2007}; \textcite{dondero_image_2009}; \textcite{dondero_semiotique_2010}; \textcite{dondero_semiotic_2014}.} The black hole's `shadow' is the consequence of the absence of light produced by its event horizon. Because of lensing effects, the shadow would appear larger than the event horizon itself. Eventually, simulations were made to show what could be expected from observations with the telescope arrangement of the EHT specifically, and EHT researchers engaged with such visualizations---they also trained, tested, and validated some of their imaging algorithms with them.\footnote{See, e.g., \textcite{bouman_extreme_2017}.}

The EHT presented its image as its primary scientific result: the image itself is supposed to serve as scientific evidence, for instance of near-horizon physics and of the structure of the black hole spacetime.\footnote{Some physicists have objected to the EHT's claims that its M87* results can serve as tests of general relativity, in part due to the outsize role that detailed astrophysical assumptions play in the simulations used in the imaging process. See \textcite{gralla_black_2019}; \textcite{gralla_can_2021}. On the interpretation of results by the EHT, see also \textcite{event_horizon_telescope_collaboration_first_2019-5} and \textcite{event_horizon_telescope_collaboration_first_2019-6}.} But if the image cannot be completely determined by the observational data, what kind of evidence is it? EHT researchers themselves observed that ``{[}r{]}econstructing an image using bispectrum measurements is an ill-posed problem, and as such there are an infinite number of possible images that explain the data. The challenge is to find an explanation that respects our prior assumptions about the `visual' universe while still satisfying the observed data.''\footnote{\textcite[916]{bouman_computational_2016}.}

Although researchers tried to justify their assumptions as rigorously as possible, they still had considerable flexibility in making various choices in the construction of images. This flexibility allowed for a range of plausible images, yet in communicating to the public, the EHT emphasized only one---which, as we will see, was an average of three images generated with different algorithms and parameters. The final image blends empirical and synthetic data in innovative and sometimes unprecedented ways.\footnote{``Synthetic data'' means data produced entirely by code, rather than by empirical observation, experiment, or measurement, such as the outputs of simulations; see \textcite[917]{bouman_computational_2016}. On ``synthetic'' in the context of data visualization in astrophysics, see \textcite{skulberg_mock_2023}.} Its construction demanded countless assumptions and choices, shaped by theoretical and methodological arguments from the physical sciences as well as considerations of aesthetics, human perception, and communication strategy. The resulting image constitutes a distinctive form of scientific evidence that historians and philosophers of science do not quite have an analytical vocabulary for yet, since even the recent literature on `data assimilation' deals with more straightforward cases---which is precisely why it should offer a highly relevant case study for epistemology.

This paper examines the epistemic and aesthetic choices involved at various stages of the construction of black hole images, focusing on the case of the EHT's iconic first image of M87*. To assist our philosophical and historical inquiry into black hole imaging research, we draw on close readings of EHT publications, technical reports, public communications, and related literature. We also compare the methodological choices and epistemological issues encountered by EHT researchers with historical and contemporary cases of other scientific projects that have grappled with related problems. Because the imaging software used by the EHT is open-source, we studied the source code and tested different options for algorithms, parameters, and colors. This enabled us to see what the images would have looked like if the EHT had made different choices. We display some alternative images in this publication, and discuss potential arguments for and against the choices that would have produced them.

Two of our figures are available online in animated and interactive form, hosted by the Niels Bohr Archive at the University of Copenhagen: an animation of an imaging algorithm constructing a black hole image step by step (Figure~\ref{fig:animated-clean}), and an interactive demonstration in which the reader can vary the parameters of another algorithm and compare the resulting images (Figure~\ref{fig:interactive-rml}).\footnote{These figures are available at \url{https://nbarchive.ku.dk/resources/coloring-black-holes/}. Their files are also archived on Zenodo at \url{https://doi.org/10.5281/zenodo.21328092}.} Since our argument concerns the variability of the plausible images that different choices produce, these figures allow the reader to directly examine some of the choices involved.

\begin{figure}[t]
    \centering
    \includegraphics[width=0.8\textwidth]{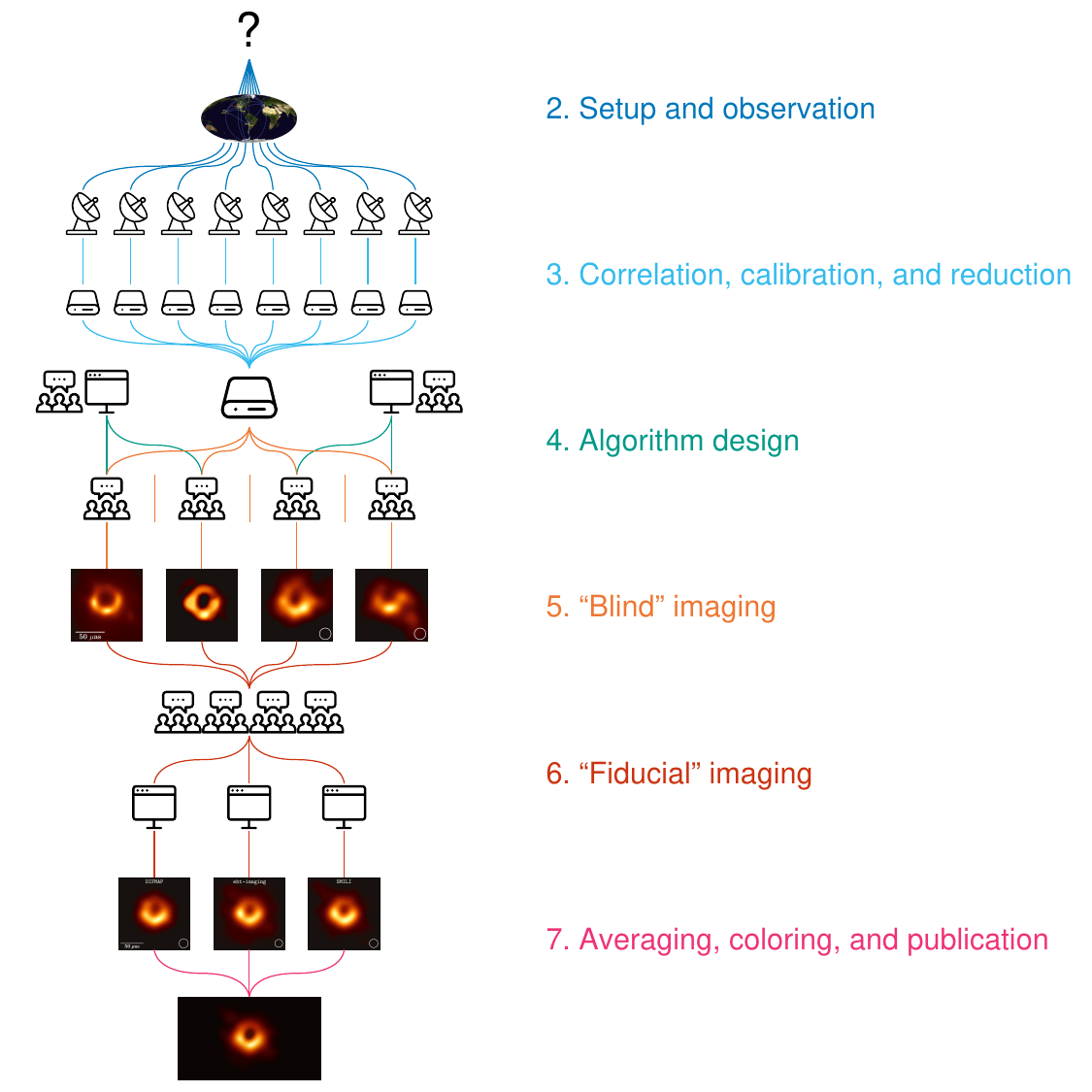}
    \caption{Diagram of different stages of data processing in the making of the EHT image of M87*. It serves as a visual index of the sections in this paper. The stages are numbered and labeled with the corresponding sections, and color-coded using the ``vibrant'' scheme by Paul Tol, which is designed for accessibility to viewers with color vision deficiency.}
    \label{fig:index-diagram}
\end{figure}

Finally, this paper features a diagram (Figure~\ref{fig:index-diagram}), which shows the different stages of data processing and serves as a visual index of the sections in the paper.\footnote{For another representation of EHT workflow, see \textcite{muhr_cartographic_2023}. Muhr highlights how the EHT's images function as epistemic ``maps'' used to navigate black hole spacetime, yielding new knowledge as they are redrawn and compared with additional data. Muhr argues that the EHT results exemplify Bruno Latour's account of `map making' and Sybille Krämer's notion of the `cartographic impulse' in the production of knowledge. We also draw the reader's attention to the illustrated and animated essay in \textcite{galison_planetary_2025}, which theorizes the EHT as offering a ``planetary vision.'' Additionally, we note that the EHT itself produced diagrams that illustrate the flow of its data: \url{https://commons.wikimedia.org/wiki/Special:Permalink/1220546000}.} The diagram is not necessarily chronological: it follows the flow of data, not of time. The data must first be recorded by radio telescopes (Section~\ref{setup}), then correlated, calibrated, and reduced (Section~\ref{correlation}), before they can be analyzed and transformed into images (Sections~\ref{algorithm}, \ref{blind}, and \ref{fiducial}), which are finally consolidated into a single image for presentation (Section~\ref{averaging}). But the work itself did not always happen sequentially. Some of it happened simultaneously, since the researchers could pursue tasks in parallel. For example, imaging researchers were developing and testing algorithms on synthetic data long before they had access to any empirical data from the telescopes. To facilitate the discussion of the epistemological issues involved at each stage, we organize both our text and our diagram based on the order of data processing rather than strict chronology. Each section in our text matches a stage in the diagram.

\section{Setup and observation} \label{setup}

\begin{figure}[t]
    \centering
    \includegraphics[width=0.8\textwidth]{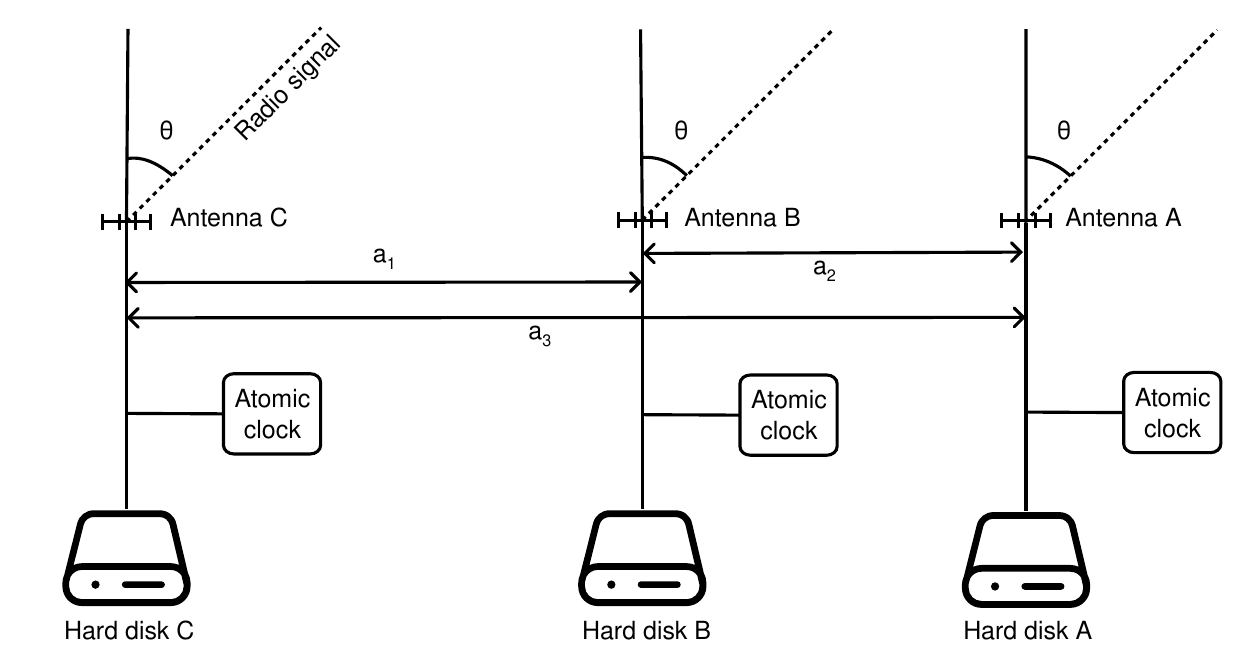}
    \caption{Diagram of a VLBI array. Incoherent signals from an extended astronomical radio source are measured at (in this diagram) three sites, and their times of arrival are accurately recorded. Using the distance between them, one can establish the correlations between the signals at the various sites. By using the van Cittert--Zernike theorem from the optics of extended sources, these measurements allow a reconstruction of the intensity distribution of the source's radiation, in other words, the image of the source. Adapted from a public-domain diagram on Wikimedia Commons.}
    \label{fig:vlbi-diagram}
\end{figure}

We start our description at the very top of our diagram, where signals from the universe hit a number of radio telescopes. Radio waves from black hole sources travel to Earth over enormous distances. In the case of the source chosen for the EHT's first image, M87*, the supermassive black hole at the center of galaxy M87, radiation has traveled some 55 million light years towards Earth. Although there are much closer astronomical objects believed to be black holes (such as Cygnus X-1, about 7,000 light years away), their masses and the diameters of their event horizons are smaller by several orders of magnitude. In terms of its observability, the size of M87* compensates for its distance. Still, to construct the image of such a source, to `image' the black hole's near-horizon shadow and emission ring, which span a distance of hundreds of astronomical units, one would need a radio telescope with an extraordinarily high resolution: the diameter of its disk would have to be larger than the Earth's radius. Yet, there is a viable alternative available: astronomers often rely on `very-long-baseline interferometry' (VLBI), which combines observations from telescopes in different locations placed very far apart (Figure~\ref{fig:vlbi-diagram}). The combining of signals from different telescopes allows one to overcome the intrinsic limitations in resolving power due to diffraction effects of the signal in the telescope.

Crucial in going from measured signals to image is the working of the \emph{van Cittert--Zernike theorem}: it states that the correlation between measurements in the plane---e.g., by two telescopes $(1, 2)$ on Earth with relative positions parametrized by $(u, v)$---is related to the intensity distribution $I$ at coordinates $(l, m)$ in the source. For two telescopes, the formula is given by:

\begin{equation}
\Gamma_{12}(u,v) = \iint I(l,m) e^{-2\pi i(ul + vm)} \, dl \, dm
\end{equation}

The formula implies that for larger relative positions $(u, v)$ between the observing telescopes a higher resolution in mapping the source can be obtained. This is a consequence of the fact that the multiple incoherently radiating points in the extended astronomical source produce a largely coherent signal when observed over large distances, due to interference effects; the more extensively one can record these correlations of the composite signal, the more accurate the reconstruction of the original source can be made. Generalizing the theorem to multiple points of measurement, instead of just two, further adds resolving power. To most accurately image an astronomical source, then, one needs as many telescopes as possible, at as large relative distances as possible.\footnote{See, e.g., \textcite{bouman_computational_2016}.}

\begin{figure}[t]
    \centering
    \includegraphics[width=0.6\textwidth]{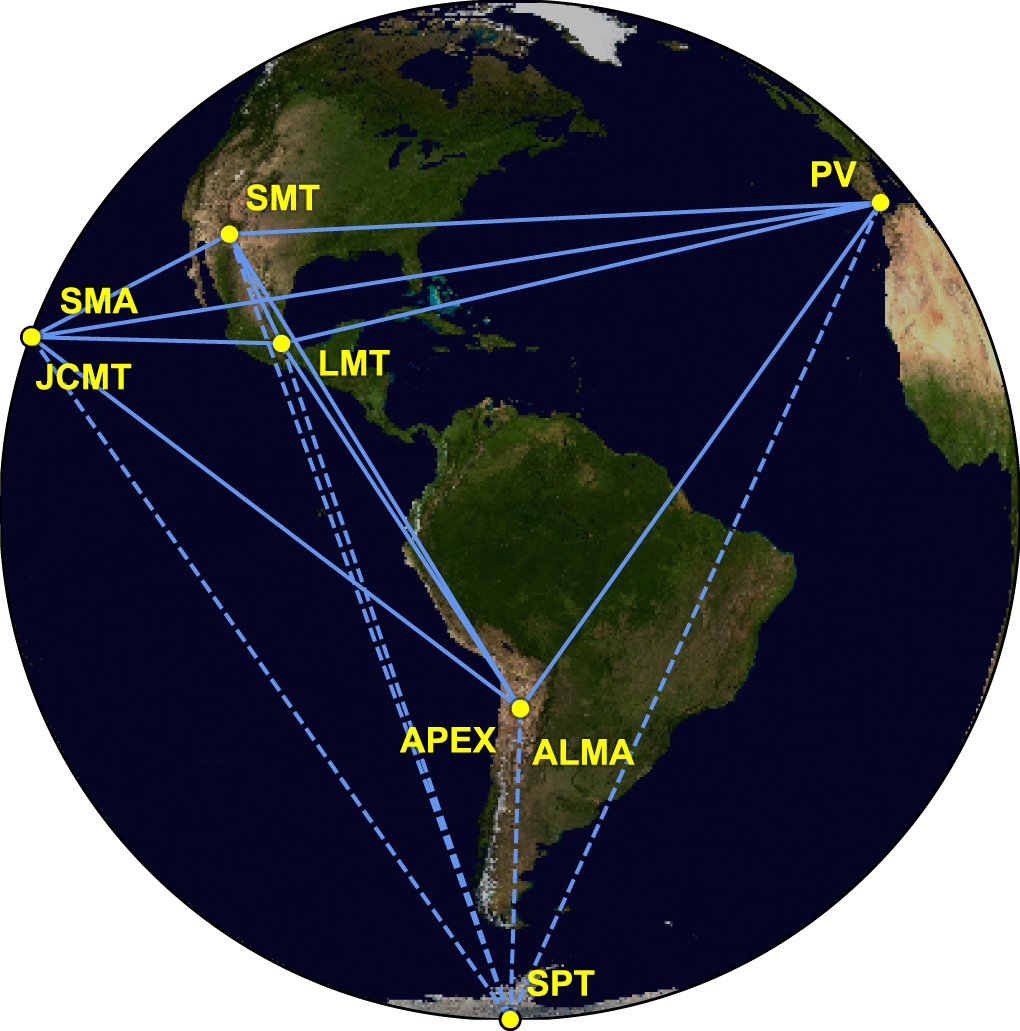}
    \caption{Eight stations of the EHT 2017 campaign over six geographic locations as viewed from the equatorial plane. Source: \textcite[3]{event_horizon_telescope_collaboration_first_2019-1}.}
    \label{fig:vlbi-earth}
\end{figure}

In 2017, the EHT used eight telescopes at very distant locations, from the Atacama Desert in Chile to Sierra Negra in Mexico to the South Pole to the Spanish Sierra Nevada to Maunakea in Hawaii (see Figure~\ref{fig:vlbi-earth}). Doing so produced several practical challenges, such as coping with weather conditions that had to be sufficiently amenable simultaneously at all locations. Furthermore, the existing equipment at some of the telescopes was not sufficient, and new equipment had to be installed. In particular, coordinated atomic clocks (hydrogen masers) were needed for synchronizing the observations at different telescopes, and because the high-resolution observations produced data at a very fast rate (64 gigabits per second), it was necessary to install computers that could support this high bandwidth for processing and storage.\footnote{\textcite{vertatschitsch_r2dbe_2015}.} It was also challenging to coordinate and maintain the equipment at the various telescopes, because they are not the same instruments, and each instrument has its own idiosyncrasies.\footnote{For a general overview of scientific and technological challenges, see the collaboration's website: \url{https://eventhorizontelescope.org/}.}

The simultaneous use of many telescopes produces challenging circumstances: it proved difficult to get all necessary telescope administrators to agree and to provide observation time to the EHT. Acquiring any observing time on these telescopes is highly competitive, especially at sites like the Atacama Large Millimeter Array (ALMA) in Chile, the largest radio telescope array in the world. Each telescope has different procedures for distributing time, is managed by different organizations, and has stakeholders in different countries. Thus, an endeavor such as the EHT produces challenges that range from the extremely intricate but mundanely logistical, all the way to the machinations of science diplomacy.\footnote{On institutional and logistical frustrations, see, e.g., \textcite[chapter 2]{skulberg_event_2021}. On international collaboration in radio astronomy, see \textcite{charbonneau_mixed_2022}; \textcite{charbonneau_mixed_2024}; \textcite{martin_radio_1978}; \textcite{mulkay_cognitive_1973}; \textcite{mulkay_methodology_1974}; \textcite{munns_single_2012}; \textcite{sullivan_early_1984}; \textcite{sullivan_cosmic_2009}; \textcite{twidle_impossible_2019}; \textcite{trangos_spatialities_2023}.} Furthermore, at several sites of the array, indigenous and environmental rights advocates had organized protests against the construction of telescopes for decades, aiming to protect the places' heritage and natural state from the invasive practices of colonial land tenure and extractivism.\footnote{At Maunakea in Hawaii, protests led by Native Hawaiian (Kānaka Maoli) activists against the planned construction of the Thirty Meter Telescope, including a roadway blockade, were active from 2014 to 2020. See \textcite{goodyear-kaopua_protectors_2017}; \textcite{hobart_at_2019}; \textcite{swanner_mountains_2013}; \textcite{swanner_instruments_2017}. On the EHT and Hawaii, see \textcite[chapter 10]{enander_facing_2025}. The EHT's reference station, ALMA in the Atacama Desert, has been the subject of conflicts between the Lickan Antay people and the Chilean state since the latter conceded lands to the observatory in 2002. In 2016, Lickan Antay activists opposed the construction of natural gas pipelines to the observatory. See \textcite{lehuede_governing_2021}; \textcite{lehuede_territories_2022}.}

We will focus on a different challenge presented by the EHT effort, addressing the various inferences involved in constructing an image from extremely noisy and sparse VLBI data. For the EHT's 2017 observation campaign that produced the image of M87*, photons from radio wave fronts (1.3\,mm wavelength) were observed during long sessions on April 5, 6, 10, and 11, 2017. On the skipped days, some telescopes suffered from poor weather or technical issues. Each telescope recorded the photons surrounding M87* by analog electrical signals, together with all the cosmic noise captured at the same time. Filtering the noise from the data, along with filling in the gaps in that sparse data through inference, became the main scientific challenge for the collaboration at the initial stage.

Around the turn of the millennium, astrophysicists Heino Falcke, Fulvio Melia, and Eric Agol had argued that it would become possible to ``image'' the shadow of the event horizon using VLBI if in the following decade a sufficient number of radio telescopes became available at sufficiently short wavelengths (at a millimeter and less: shorter wavelengths improve the reliability of the van Cittert--Zernike theorem for mapping a source).\footnote{\textcite{falcke_viewing_2000}.} Figure~\ref{fig:m87-public} was produced on the basis of observations at 1.3\,mm. More recently, detections of multiple sources on the wavelength of 0.87\,mm have been announced, which are promising for future EHT observations.\footnote{\textcite{raymond_first_2024}.}

Generally, computer code and human input are now essential to complement the sparsely available data and to reliably guide the mapping---provided by the van Cittert--Zernike theorem---from data to `reconstructed' images of the source. Due to the reliance on algorithms for, e.g., signal and noise selection, benchmarking of priors in Bayesian methods, or the production of ``synthetic data,''\footnote{\textcite[917]{bouman_computational_2016}.} any presumed dividing line between digital signal processing software and the material instrument of the telescope has become entirely blurred. The algorithms, in the words of computer scientist and EHT member Katie Bouman, are ``part of the telescope.''\footnote{Bouman cited in \textcite[23]{doboszewski_robustness_2024}. In the context of the EHT, \textcite[23]{doboszewski_robustness_2024} have argued for treating imaging algorithms ``analogously to instruments'' where the different algorithms allow for ``varying the experiment, even though significant variation in the physical array is not viable.'' Another recent suggestion for expanding the notion of an experiment to include the laboratory's environment has come from \textcite{nichols_hidden_2024} and \textcite{nichols_constructing_2022} on LIGO.}

\section{Correlation, calibration, and reduction} \label{correlation}

We are now at the second stage in our diagram: the first round of data processing. What did it take for the EHT to gain confidence in the data it used to image a black hole shadow for the first time? As part of the data preparation, `correlation' was used to bring together data from different telescopes such that the times the signals were received at different telescope sites matched. The `calibration' of data also played a crucial role in laying the ground for imaging and analysis by correcting for errors in the data, and taking into account instrument particulars and weather conditions at different sites of observation. This stage entailed a great amount of work, particularly for postdocs and PhD students. In this section, we describe the important stages of correlation and calibration of telescope data, as the EHT moved from signals from observation to data that could be used to make images. We also address the discussions of what the term `raw data' entailed following the publication of results by the EHT.

\begin{figure}[t]
    \centering
    \includegraphics[width=0.6\textwidth]{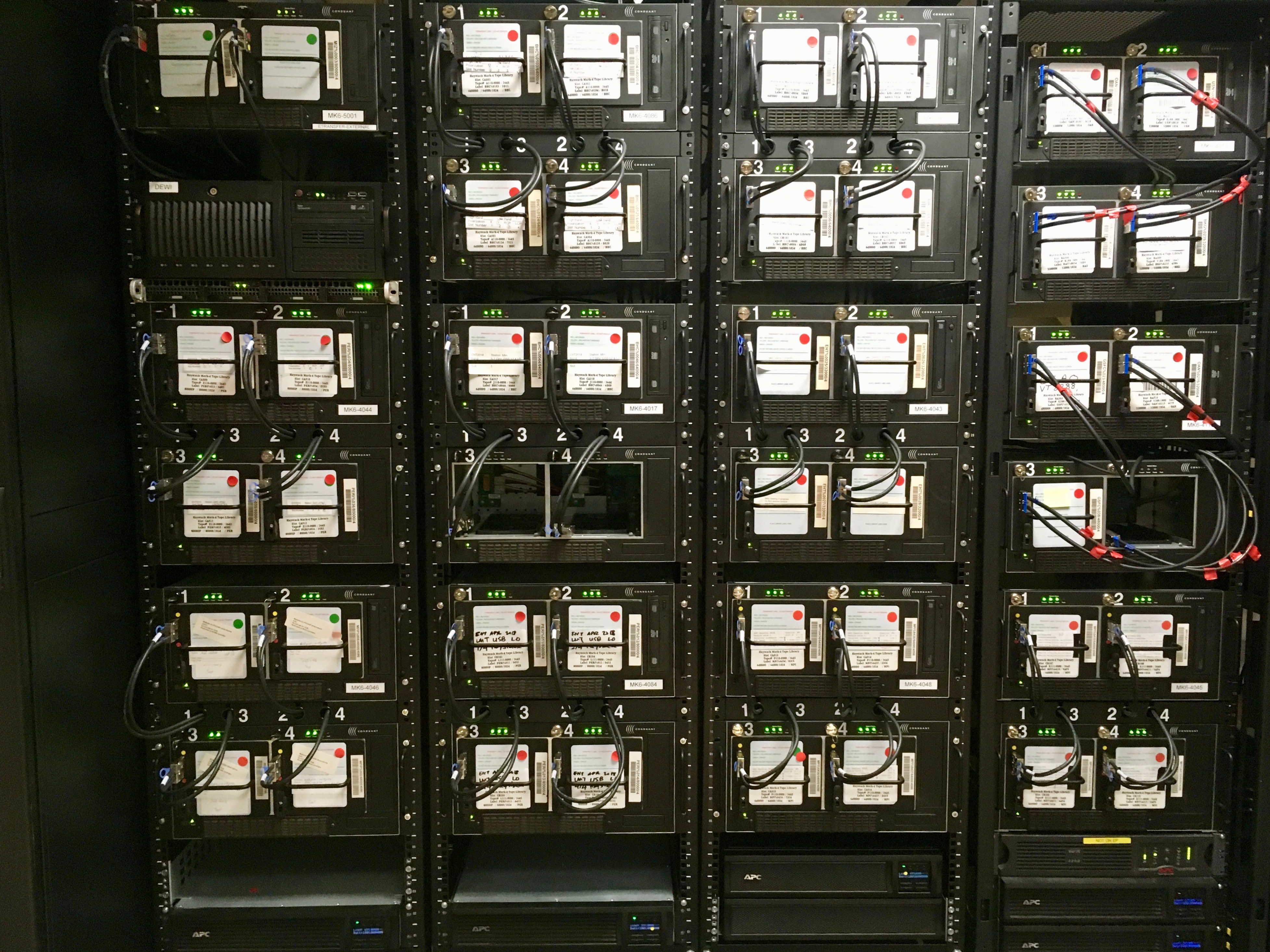}
    \caption{Correlator at the MIT Haystack Observatory, 2018. Photo: Emilie Skulberg.}
    \label{fig:correlator-photo}
\end{figure}

We saw in the previous section that the radio signals travel from the light surrounding the black hole to the telescopes. At the telescopes, the radio signals become raw voltages that get converted into digital data. These are time stamped and stored in hard disks. To operate under extreme conditions of temperature and pressure, for example at the South Pole, the hardware needs to be designed specifically for those conditions: such hard disks are filled with helium and hermetically sealed, and the cables are thermally insulated.\footnote{\textcite{event_horizon_telescope_collaboration_first_2019-2}; \textcite[8, 19, 22]{event_horizon_telescope_collaboration_first_2022}.} Because the amount of recorded data was very large, on the scale of petabytes, it would have taken too long to transfer the data over the internet; instead, the physical hard disks themselves were transported via air travel.\footnote{\textcite{event_horizon_telescope_collaboration_first_2019-2}; \textcite{goddi_first_2019}.} These were taken to two physical facilities, one at the MIT Haystack Observatory in Westford, Massachusetts (Figure~\ref{fig:correlator-photo}), and another at the Max Planck Institute for Radio Astronomy in Bonn, Germany.

Radiation recorded at individual telescopes was synchronized (i.e., matched to a single time coordinate) by aligning the local atomic clocks using GPS. In the `correlation' stage, delays due to the Earth's rotation and to varying atmospheric conditions at the different sites were also corrected for. Typically, such corrections are based on a priori models, for example of the Earth's geometry. Since the van Cittert--Zernike theorem uses correlations as input, this stage is essential to make raw data into possible signals from which noise can be filtered.

The correlation stage involves not only specialized software but also, in some cases, specialized hardware. From the earliest days of radio astronomy, the correlation of interferometric signals typically relied on computers (known as `correlators') designed specifically for this purpose. More recently, such `application-specific integrated circuits' have been increasingly replaced by software that runs on general-purpose computing clusters. The most commonly used `software correlator' is known as DiFX.\footnote{\textcite{deller_difx_2007}.} The EHT ran DiFX on computing clusters in Westford and Bonn.\footnote{This process of correlation may be seen as an instance of what historian of science Paul Edwards, in his study of climate modeling, has called ``making data global'': building coherent global data sets from heterogeneous sources at disparate locations (in our case, radio telescopes with varying designs, resolutions, and idiosyncrasies); see \textcite[chapter 10]{edwards_vast_2010}.}

Next came the laborious stage of calibration and reduction. Its aim was to decrease the amount of data and to make it easier to work with for analysis. This involves correcting for errors due to the idiosyncrasies of the different telescopes, standardizing the data, and correcting for errors tied to other instrumentation, atmospheric effects, and delays caused by the different moments of arrival of the signals at each telescope site. To calibrate the data, radio astronomers rely on a combination of strategies. One strategy is to use the most trusted instrument as a reference for the others. In the case of the EHT, the reference station was ALMA, in part because of its higher sensitivity. Another strategy is to point the instruments to the `cold sky,' or to a compact source that has a precisely known location and a familiar and strong brightness. Such sources are often planets or quasars: among others, the EHT used the quasar 3C 279, because it is a strong radio source with a clear nuclear core.

Yet another strategy is to point the instruments to artificial sources, such as a known standardized black-body radiation source. In the case of the EHT, a sheet of polyurethane foam, designed to absorb radio waves and sold commercially as Eccosorb, was put in front of the receiver for this purpose. By comparing measurements of thermal radiation from this opaque sheet and from the sky, the researchers were able to estimate the sensitivity of their instrument. To obtain rapidly alternating measurements of the sky and of the opaque sheet, the researchers used the so-called `chopper' method, which is common in radio astronomy and was introduced in the 1970s. At the time, astronomers found that repurposing a cheap, commonplace instrument---a two-bladed chopper driven by a synchronous motor---in front of the receiver worked conveniently well for this purpose. ``The extreme simplicity and reliability of this technique has resulted in its adoption for most new millimeter-wave radio astronomy systems, even at the longer-millimeter wavelengths,'' two researchers wrote in 1973.\footnote{\textcite[64]{penzias_millimeter-wavelength_1973}.}

Although there are qualitatively different kinds of noise and error, EHT researchers chose a pragmatic approach to manage them. They quantified the errors and kept track of an error ``budget.''\footnote{See, e.g., \textcite[Table 3]{event_horizon_telescope_collaboration_first_2019-3}. For a discussion of calibration and uncertainty budgeting in scientific measurements, see \textcite{tal_calibration_2017}.} When a particular kind of error was too difficult to remove or mitigate, the researchers could choose to ignore it if their budget could afford it. For example, when they found a ``small glitch in the ALMA correlator,'' they decided that the amplitude loss from the glitch was less than 0.1\%, so they chose simply to leave this error unaddressed.\footnote{\textcite[Appendix A.2]{event_horizon_telescope_collaboration_first_2019-3}.} At the same time, they tried to get an understanding of anything that appeared `odd' in the data, checking in particular aspects of the data that did not match their knowledge about the source or the instruments.

The standards of calibration were first used on ``calibration targets'' that were known sources, such as the quasar 3C 279. Only after performing the analysis on calibration targets did the researchers proceed to imaging and analyzing black hole candidates. The EHT researchers used three independent ``pipelines'' for calibration, that is, three scripts for processing the data using different software packages, called HOPS, CASA, and AIPS. ``Following data validation and pipeline comparisons,'' the EHT authors wrote, ``a single pipeline output was designated as the primary data set of the first EHT science data release and used for subsequent results, while the outputs of the other two pipelines offer supporting validation data sets.''\footnote{\textcite[4]{event_horizon_telescope_collaboration_first_2019-1}.} In their communication of calibration, the EHT emphasized that the process was automated and executed with independent software packages.

Before correlation and calibration, the EHT gathered a massive amount of data, on the scale of petabytes. This meant that publishing the full data would be impractical, the collaboration found. When the EHT published the M87* image and papers in April 2019, it released a small data set of a few megabytes, which was already calibrated.\footnote{\textcite{event_horizon_telescope_collaboration_dataset_2019}.} This data set was subsequently reanalyzed by multiple groups outside the EHT.\footnote{\textcite{arras_variable_2022}; \textcite{carilli_hybrid_2022}; \textcite{lockhart_how_2022}; \textcite{miyoshi_jet_2024}; \textcite{patel_reproducibility_2022}.} One of these---consisting of Makoto Miyoshi, Yoshiaki Kato, and Junichiro Makino---then objected that this limited data set did not allow for a proper independent verification of the EHT's results, not only because it compressed a wide frequency band into a single channel (a choice they deemed ``extremely rare and unsuitable for public data'') but also because it lacked crucial information about correlation and calibration.\footnote{\textcite[36]{miyoshi_jet_2022}.} Could the EHT somehow release the ``raw data''?

In May 2022, the EHT released larger data sets, on the scale of gigabytes and terabytes.\footnote{\textcite{event_horizon_telescope_collaboration_dataset_2022}.} These included what the EHT called ``L1'' (Level 1) data products. This level mainly comprises ``validated correlation products, and an associated metadata package containing information on telescope sensitivities and other parameters necessary for the flux density calibration of the correlated visibilities''---relatively but not absolutely raw data, after one first step of processing.\footnote{\textcite[3]{koay_metadata_2023}.} The EHT distinguished this from ``L2'' (Level 2) data produced by additional calibration steps. By calling the data products ``L1,'' the EHT suggested that while the published data were not completely unprocessed (which would be called ``L0'' in other fields of science), their processing was minimal.\footnote{This terminology of ``levels'' of data processing is not unique to the EHT, but used across astronomy and in various other fields of science. For example, NASA categorizes its data of the Earth's geophysical properties from Level 0 (``unprocessed'' data at full resolution) to Level 4 (results from models or analyses of processed data). See \textcite{nasa_data_2025}; \textcite[13]{king_eos_2004}.}

This second release was not enough to satisfy the critics, however, who continued to call for less processed, higher-resolution ``raw data'' from the EHT's 2017 observations. The critics published another reanalysis, this time focused on the EHT's later image of a different black hole (Sagittarius A*). The EHT had collected data of both M87* and Sagittarius A* during the same observing campaign in 2017. The critics wrote: ``Checking the raw output data of the correlator is necessary to identify the cause of discrepancies in the closure quantity. If the recorded data are available, the correlation process should be reproduced.''\footnote{\textcite[3247]{miyoshi_independent_2024}.}

The EHT replied that its second release did actually offer ``raw data,'' stating: ``The EHTC welcomes critical, independent analysis and interpretation of our published results. We publish detailed descriptions of our methods as well as raw data, data products, and analysis scripts to facilitate transparency, rigor, and reproducibility.'' Yet, the EHT also argued that the critics' paper ``erroneously'' claimed that the EHT had not been fully forthcoming: ``all of 2017 EHTC raw and calibrated data have been publicly available since May 2022.''\footnote{\textcite{event_horizon_telescope_collaboration_response_2024}.} The EHT and its critics thus disagreed over what may count as ``raw data.''

As information scholar Geoffrey Bowker and media historian Lisa Gitelman have put it, ``\,`raw data' is an oxymoron.''\footnote{\textcite{gitelman_raw_2013}. See also \textcite[184]{bowker_memory_2008}.} Scientists still use the notion for practical purposes, for instance when aiming to facilitate efforts at scientific reproducibility. Philosopher of science Sabina Leonelli has emphasized the different ways `raw data' are conserved and ``packaged'' (e.g., format and metadata), and how these ways can vary between areas of research.\footnote{\textcite[89--90]{leonelli_data-centric_2016}. On differing views of raw data in meteorology and climate science, see also \textcite[chapter 11]{edwards_vast_2010}.} 
The controversy surrounding raw EHT data demonstrates how the meaning of `raw data' can be contested even among professional peers, exactly in the context of reproducibility: even when scientists agree that the public availability of `raw data' is important for reproducibility, they may disagree about what data count as sufficiently unprocessed to enable \emph{independent} verification. In the case of the EHT, this issue is especially pronounced because data processing is seen by some EHT researchers as part of the instrument itself.\footnote{\textcite[23]{doboszewski_robustness_2024}.}

In its paper on correlation and calibration (Paper III) from the 2019 release, the EHT had reassured the reader of its thorough data preparation: ``Given the uniqueness of the data set and scientific goal of the EHT observations, our processing focuses on the use of unbiased automated procedures, reproducibility, and extensive review and cross-validation.''\footnote{\textcite[2]{event_horizon_telescope_collaboration_first_2019-3}. See also \textcite{blackburn_eht-hops_2019}; \textcite{janssen_rpicard_2019}. When the EHT released their results in 2019, they published six papers on different aspects of the results, moving from an overview of the results (Paper I), observation and instrumentation (Paper II), data processing and calibration (Paper III), imaging (Paper IV), the physical implications of the observed ring (Paper V), and the interpretation of results focusing on the mass and shadow of M87* (Paper VI). See \textcite{event_horizon_telescope_collaboration_first_2019-1}; \textcite{event_horizon_telescope_collaboration_first_2019-2}; \textcite{event_horizon_telescope_collaboration_first_2019-3}; \textcite{event_horizon_telescope_collaboration_first_2019-4}; \textcite{event_horizon_telescope_collaboration_first_2019-5}; \textcite{event_horizon_telescope_collaboration_first_2019-6}.} The EHT emphasized the objective quality of its data by highlighting the mechanical nature of the initial processing\footnote{On the historical relation between `objectivity' and mechanical processing of data, see \textcite{daston_objectivity_2007}. For an interpretation of different kinds of objectivity at different stages of EHT image production, see \textcite{galison_governing_2023}.} and by further pointing out that it was aiming to provide thoroughly vetted data, readied for result reproduction efforts. We will see in Section~\ref{algorithm} how researchers in the EHT tested the imaging algorithms and their potential biases in multiple ways. When describing the imaging stage in peer-reviewed papers, the EHT stressed that while bias was unavoidable, they had taken several steps to ensure the validity of their results.

\section{Algorithm design} \label{algorithm}

How to convert the correlated and calibrated telescope data into an image? Radio astronomers call this problem ``image synthesis'' or ``image reconstruction,''\footnote{See, e.g., \textcite{thompson_interferometry_2017}.} and it is found in the third step in our diagram. Note the prefix ``re-'' in the word ``reconstruction,'' which suggests an epistemic commitment to an existing image that needs to be assembled again. Although EHT researchers have to construct an image from limited available data, they often frame their task as the reconstruction of an imagined, ``true image'' that reflects how the source would ideally look in a hypothetical scenario free of noise or error. In this section, we will look at the design of the algorithms that `synthesize' or `reconstruct' images in this sense, focusing on the assumptions made by different algorithmic techniques.

In radio astronomy, the telescopes do not directly measure the sky brightness distribution. Rather, as we saw, they measure the interference patterns between the frequencies picked up in pairs of antennas. The resulting measurements are called \emph{visibilities}. The sky brightness distribution needs to be inferred from such visibilities. Under certain ideal conditions, the sky brightness distribution would be equal to the inverse Fourier transform of the visibilities.\footnote{A Fourier transform is a mathematical operation that converts a signal into a sum of frequencies; the inverse transform recovers the signal from the frequencies.} This mathematical relationship between the visibilities and sky brightness is based on the van Cittert--Zernike theorem (see Section~\ref{setup}), which makes several assumptions, such as the incoherence of the extended source, a very long distance to the source, and homogeneity of the medium through which light travels. Because the measurements are available only for an incomplete sample of signals, the application of an inverse Fourier transform to the visibilities produces a ``dirty image'' of the source with multiple flaws, including sidelobes. There are several methods for producing a ``clean image'' from a ``dirty'' one, each of which rests on different assumptions. To image M87*, EHT researchers used two classes of methods. As we will see in later sections, the final image was the average of images produced using both classes of methods.

\begin{figure}[t]
    \centering
    \includegraphics[width=\textwidth]{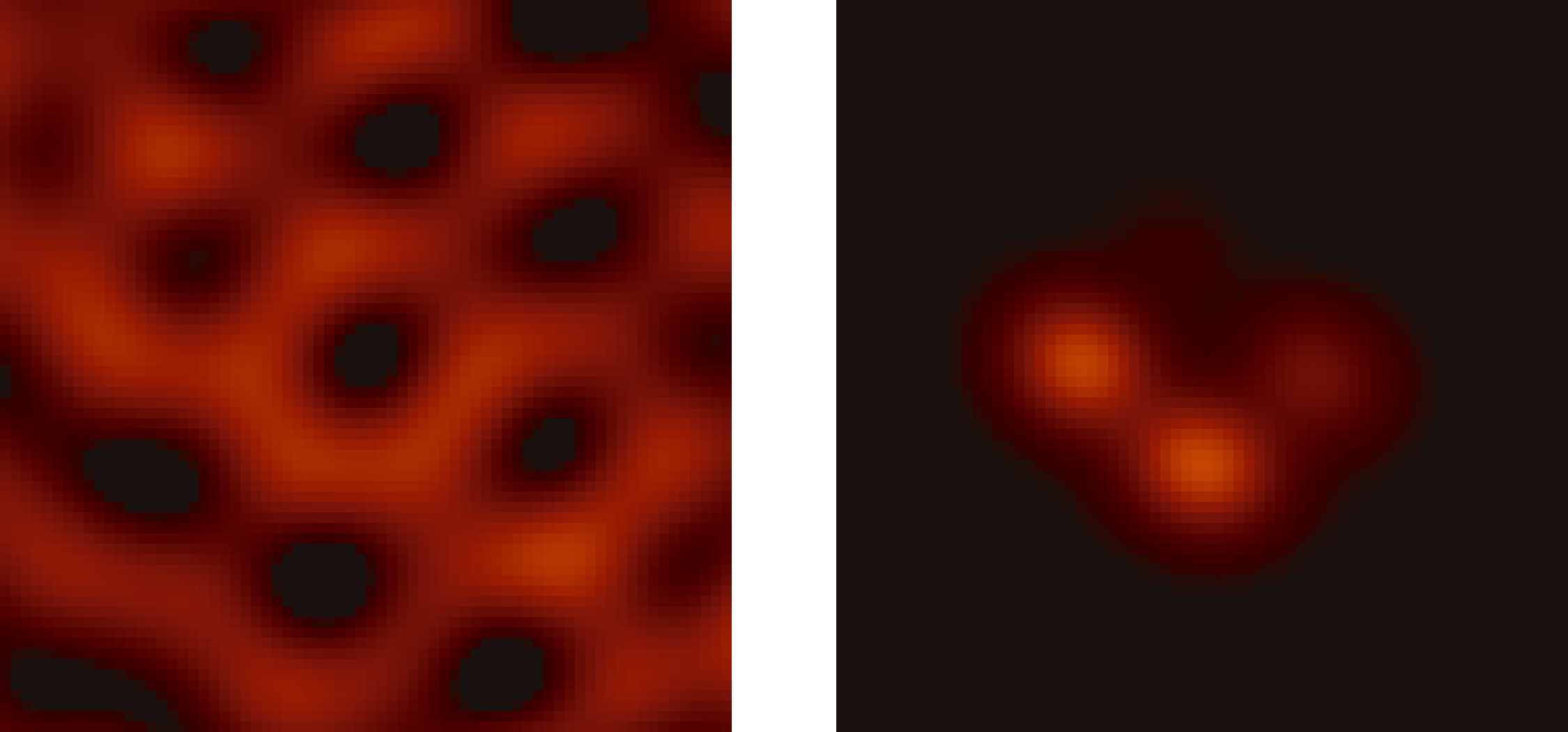}
    \caption{Preview of an animated demonstration of the CLEAN algorithm, based on the EHT's script for the Difmap software. The animation shows the ``dirty'' or ``residual'' image (left) and the ``clean'' image (right) of M87* at successive steps of the algorithm, spanning 1,300 iterations. This still frame shows an early step, after 100 iterations, and illustrates how the algorithm represents the sky as a collection of point sources. The full animation is available at \url{https://nbarchive.ku.dk/resources/coloring-black-holes/} and archived at \url{https://doi.org/10.5281/zenodo.21328092}.}
    \label{fig:animated-clean}
\end{figure}

The first class consisted of methods derived from the CLEAN algorithm, originally published in 1974 by astronomer Jan H{\"o}gbom.\footnote{\textcite{hogbom_aperture_1974}.} This algorithm and its derivatives are the most conventional and established class of methods for `cleaning' images in radio astronomy. CLEAN represents the sky as a collection of point sources in an otherwise empty field, and requires an assumed model of how the telescope array responds to a point source (called the ``dirty beam''). CLEAN looks for the brightest point in the dirty image, assumes that this brightest point is a source, subtracts a dirty beam from the same location, and repeats this process with the next brightest point until there are no sufficiently bright points left in the ``residual image'' (or until some other stopping criterion is reached). Each point source is added as a ``clean beam,'' which has an idealized mathematical shape (usually a normal distribution), to an initially empty image. Thus, the clean image is not produced by simply subtracting noise from the dirty image. Rather, it is produced by adding idealized points to an initially empty image. The location of the points is determined by iteratively finding and subtracting the brightest points from the dirty image (see the animated version of Figure~\ref{fig:animated-clean}). In this sense, the metaphor of `cleaning' can be slightly misleading, since it suggests the mere removal of `dirt' from the original image rather than the production of a new image on an empty field.\footnote{On cleaning as a normatively and ontologically consequential practice in physics experiments rather than a mere removal of contamination, see \textcite{deswart_cleaning_2026} on the XENONnT dark matter detector.}

Because CLEAN represents the sky as a collection of point sources, it tends to break up extended features into multiple small ones.\footnote{See \textcite[77]{bouman_extreme_2017}.} For sources with extended emissions, such as black hole shadows, this is a key limitation of CLEAN. Moreover, in the case of the EHT, the use of CLEAN required an additional process of ``self-calibration'' in order to account for phase errors due to the Earth's atmosphere, as well as differences in amplitude calibration between different telescopes.

Running CLEAN is not a fully automated, one-click affair. In practice, researchers use interactive programs that allow them to inspect and manipulate the data at various stages and iterations of the algorithm. For the M87* image, the EHT used the program Difmap, which was developed at Caltech in the 1990s. The use of such interactive programs involves expert judgment and skillful manipulation of the data. In an effort to make the process more transparent and reproducible, EHT researchers have advanced a ``scripted'' approach to the use of CLEAN, in which the data manipulation is written on a computer script that can be read and run by others.\footnote{\textcite[12]{event_horizon_telescope_collaboration_first_2019-4}.}

The second, newer class of methods, known as ``regularized maximum likelihood'' (RML), frames the imaging task as an optimization problem, typically interpreted within a Bayesian statistical framework. The task is to construct an image $I$ that maximizes the \emph{likelihood} $P(V \mid I)$---the probability that the observed visibilities $V$ would have been produced given that $I$ is the true source. This likelihood is computed from a probabilistic model of how multiple sources of noise and error (including atmospheric phase, amplitude calibration, thermal noise, and gain error) affect the visibilities. These sources of noise and error are typically assumed to follow normal distributions. An algorithm generates many possible images, then computes the probability that each image would have produced the given visibilities according to the model. The goal is to select the `optimal' image, for which this probability is as high as possible.

Note that the optimal image is not necessarily likely to resemble the source. This is because the likelihood $P(V \mid I)$ and the \emph{posterior} $P(I \mid V)$---the probability that the image $I$ is the true source given the observed visibilities $V$---are distinct quantities. To put it another way, there is a difference between the following two questions:

\begin{enumerate}
\item \emph{Likelihood}: If this image were true, what is the probability that these visibilities would be observed?
\item \emph{Posterior}: If these visibilities were observed, what is the probability that this image is true (or sufficiently similar to the true one, in general shape or in particular details)?
\end{enumerate}

A high likelihood does not by itself entail a high posterior probability. Bridging the gap between the two requires substantive epistemic commitments, such as judgments about an image's consistency with different imaging methods. One way to narrow the gap is to consider an ensemble of images that fit the data well (instead of only the single best-fit image).\footnote{EHT researchers have since developed methods for analyzing an ensemble of images drawn from the posterior probability distribution. See, e.g., \textcite{broderick_themis_2020}; \textcite{pesce_d-term_2021}.}

The likelihood--posterior gap is especially wide in this case because the EHT's sparse baseline coverage leaves the image radically underdetermined: many possible images fit the visibility data equally well. Researchers must then choose among them. Such a choice relies on theoretical expectations about the source, incorporated into the optimization problem as additional ``regularization'' terms that function, in Bayesian terms, as implicit priors on image structure. For example, the image can be `regularized' in ways that favor smoother images (``total squared variation regularization'') or images with a smaller total number of bright pixels (``$\ell_1$ norm regularization''). The image can also be regularized based on pixel-to-pixel similarity with an assumed ``prior image'' (``maximum entropy regularization''). For the image of M87*, the EHT combined a variety of regularizers, with different weights for each. A particularly important regularizer favored pixel-to-pixel similarity to a circular Gaussian distribution, i.e., it penalized images with bright pixels farther from the center. One justification for this regularizer is that other methods such as CLEAN, which do not assume a circular Gaussian prior, also produced images with a circular shape. Still, one could argue that this regularizer makes a heavy-handed assumption about the shape of the image. Such difficult choices illustrate a larger point: it is not the case that the researchers simply observed an image that matched and confirmed their expectations about the shape of the black hole. Rather, certain expectations were built into the algorithms which generated the image in the first place.

The EHT produced images using both CLEAN and RML methods. This mixture of methods was the EHT's response to a double bind. On the one hand, because the EHT sought to make an image of a novel object, sticking to a time-tested method such as CLEAN would afford greater confidence in the result. On the other hand, the more established methods suffered from limitations when applied to this novel object. CLEAN not only required phase and amplitude calibration but also represented the sky as a collection of points, which did not work well for sources with extended emissions. Researchers had to develop new RML methods, algorithms, and software packages especially for the EHT. The EHT's reliance on new, custom-made methods amplified worries of imaging artifacts and potential doubts in the reception of the result by other scientists. The EHT sought to resolve this double bind by combining older and newer methods: by demonstrating that both classes of methods produced results with a consistent structure, and by using results from more established methods to justify assumptions built into less established ones.

The imaging process happened in two main stages, called ``blind'' and ``fiducial'' imaging. At the ``blind'' stage (discussed next in Section~\ref{blind}), the EHT Imaging Group was divided into four independent teams, who were restricted in their communication in order to cross-check each other's results. Two of the teams used CLEAN methods and two used RML. Each team was free to manipulate its algorithms and parameters as it wished, tweaking and optimizing the results by trial and error. After the results of each team were revealed to all, the Imaging Group proceeded to the ``fiducial'' stage (discussed in Section~\ref{fiducial}), which produced new images by running the algorithms again, but with parameters chosen in a less ad hoc, more mechanical way. The fiducial stage was divided into three pipelines, two of which used RML methods, and the other used CLEAN. The three resulting images were then averaged to create a final image, as we will explain in Section~\ref{averaging}.

\section{``Blind'' imaging} \label{blind}

When is it justified to have faith in an image of an object which has never before been `seen'? This question was central to researchers in the EHT Imaging Group throughout the process of making images with their algorithms. Such epistemic concerns about imaging the shadow of a black hole for the first time were addressed in various ways and at different stages. During the training for image making, and the early stage of image production (the later stage will be described next in Section~\ref{fiducial}), researchers focused on hiding images from view and revealing them at particular points. The preparation and the early stage of imaging were focused on what was \emph{seen} or not, with the aim of avoiding `bias.' By setting up independent pipelines of image production, EHT researchers thought, they could take a first step towards faithful images.

One of the motivations for the EHT `Imaging Group' to carefully test its own work was that it wanted to avoid a scenario like that of the `BICEP2' observations: initially, it had been claimed that the BICEP2 instrument had detected primordial gravitational waves, but the signal turned out to be caused by cosmic dust. As we will see, the EHT integrated many precautionary measures into its imaging process: it tested its methods in multiple ways and at different stages. Members of the Imaging Group wanted to avoid circularity and bias: they wished to avoid reproducing with their imaging algorithms what they already expected to see, and they elaborated on these points in their journal publications to reassure readers.

\begin{figure}[t]
    \centering
    \includegraphics[width=0.4\textwidth]{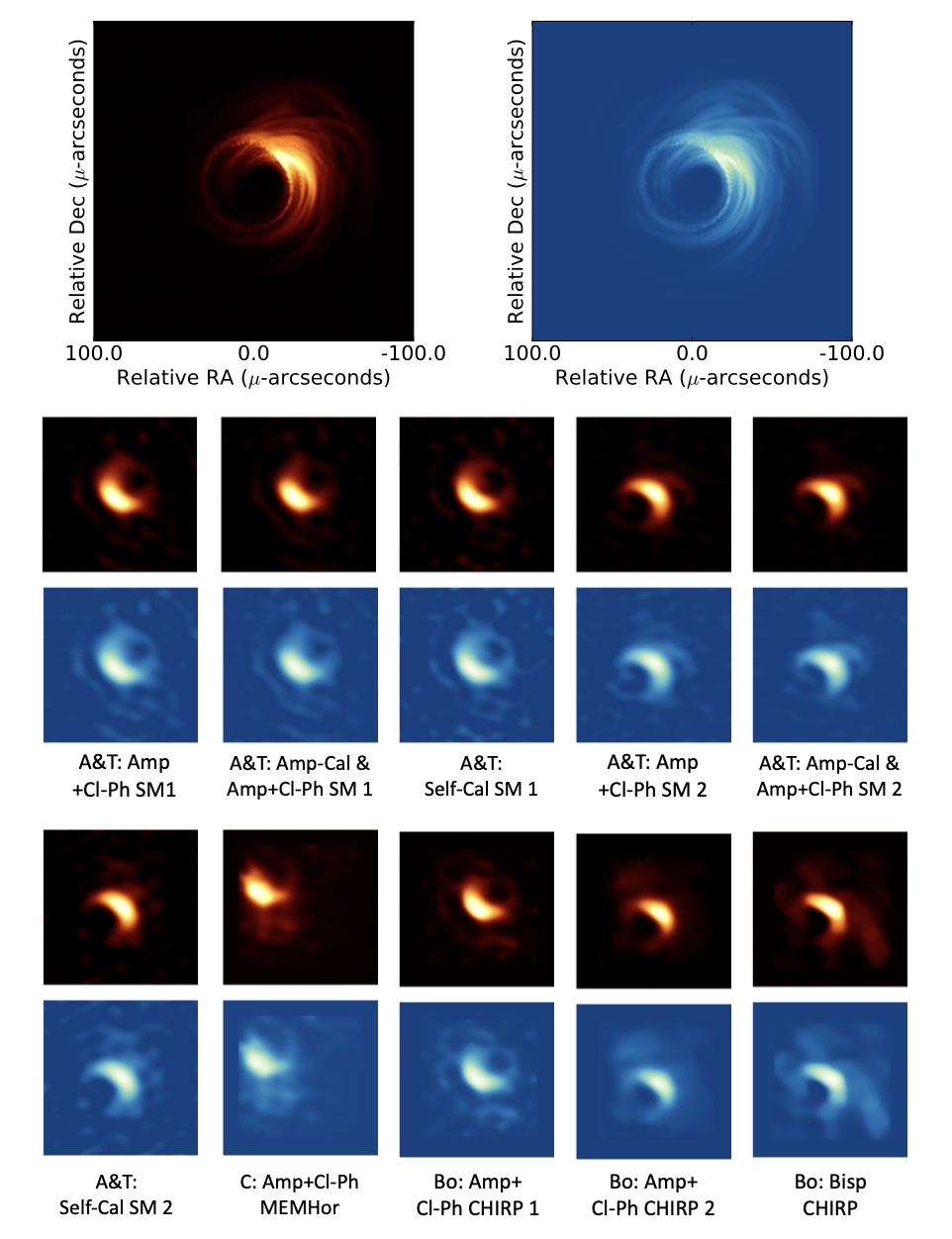}
    \caption{Example of image reconstructions in an Imaging Challenge. In this data set, the original ``truth image'' (above, shown in two color maps) was a black hole simulation by Hotaka Shiokawa; see  \textcite{shiokawa_general-relativistic_2013}. The smaller images below are image reconstructions. Columns of image reconstructions in two color maps were shown, with the creators and methods listed below. For example, the first column from the left was made by Kazu Akiyama and Fumie Tazaki who used a ``sparse modeling'' method. Source: \textcite[160]{bouman_extreme_2017}. Courtesy of Katherine L. Bouman.}
    \label{fig:grmhd-reconstructions}
\end{figure}

The development of new imaging algorithms began before the data used for the black hole images had been obtained, when members of the Imaging Group trained themselves in making images of various (not necessarily astronomical) objects that were hidden from them. In 2016, while still a PhD candidate at MIT, Katie Bouman made a website that was used as part of the organization of what the group called ``Imaging Challenges.''\footnote{\textcite{bouman_vlbi_2016}.} Bouman came to play an important role in how the training of the Imaging Group took shape. Using the website, researchers worked with ``test data'' to prepare themselves for imaging with actual black hole data. Artificial test data were typically extracted from visualizations made by simulations of black holes (e.g., Figure~\ref{fig:grmhd-reconstructions}, reproduced in Bouman's PhD thesis). The researchers then had to recreate these visualizations with their imaging algorithms from the test data (e.g., the bottom two rows of Figure~\ref{fig:grmhd-reconstructions}). In this way, the tradition of visualizations made using simulations of black holes, which in theoretical physics and computational astrophysics can be traced back to the late 1970s, became an integral part of image production in radio astronomy.\footnote{In 1978, two visualizations were produced. One was a film clip made by Leigh Palmer, Maurice Pryce, and William Unruh at Simon Fraser University. The film clip was shared during lectures but it was not published. See, however, the later publication \textcite{matzner_grand_1985}. Luminet's image was first published in \textcite{carter_trous_1978}. It was then published in \textcite{luminet_image_1979}.} The researchers used not only traditional algorithms based on CLEAN but also newer algorithms custom-made by members of the EHT's Imaging Group.

\begin{figure}[t]
    \centering
    \includegraphics[width=0.4\textwidth]{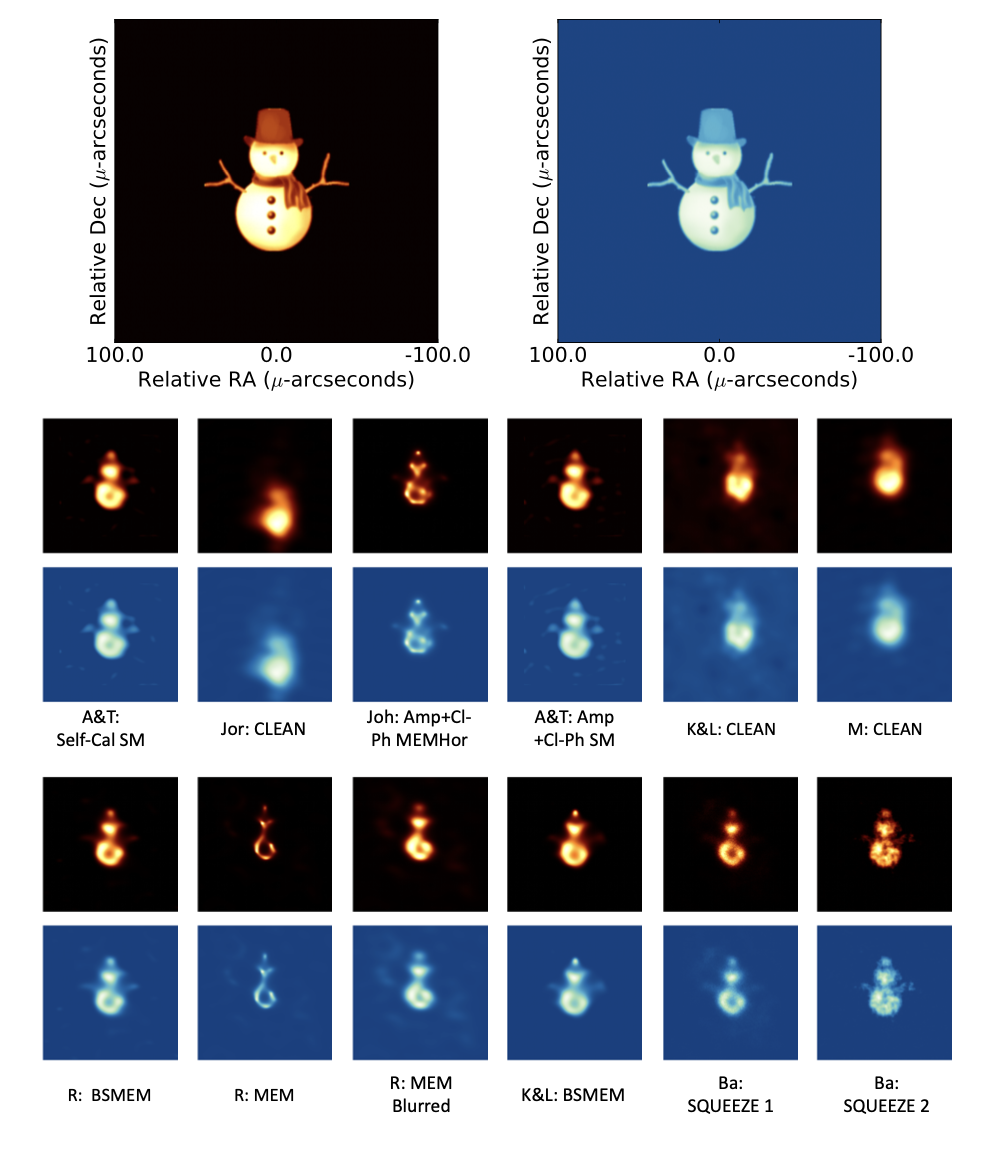}
    \caption{The top image of a snowman is the ``truth image'' in two color maps, and the reconstructions of the image are below (as in Figure~\ref{fig:grmhd-reconstructions}). Source: \textcite[164]{bouman_extreme_2017}. Courtesy of Katherine L. Bouman.}
    \label{fig:snowman-reconstructions}
\end{figure}

Importantly, participants in the Imaging Challenges did not know what they were making images of. In fact, in these Imaging Challenges they ended up imaging anything from black hole depictions (based on simulations) to a snowman (Figure~\ref{fig:snowman-reconstructions}). ``Judges,'' typically senior members of the EHT, then had to look at the image reconstructions to guess what they depicted.

While training on the basis of simulated images, members of the Imaging Group developed concerns about how their preconceptions of what black holes and their shadows looked like might influence any end result. Since no one had ever observed a black hole shadow before, how could members of the Imaging Group know whether their images were accurate, or whether their a priori ideas of what a black hole looked like might have influenced their images in ways they should not have? In the early stage of imaging, the language used by the group when expressing the need to address this concern was that one needed to suppress `bias.' Judges were also focused not only on how good the reconstructions seemed to be or on what they depicted, but also on how similar or different the images were from each other, as a measure for the reliability of the collaboration's procedures.

In several synthetic data sets that were part of the challenges, general-relativistic magnetohydrodynamic (GRMHD) simulations of black holes were used. GRMHD simulations are important tools for exploring the dynamics of magnetized gas around a black hole. Commenting on the reconstructions of a data set showing the shadow of a black hole in a GRMHD simulation by Hotaka Shiokawa (Figure~\ref{fig:grmhd-reconstructions}),\footnote{\textcite{shiokawa_general-relativistic_2013}.} one of the judges commented: ``Overall a decent batch of reconstruction, but what is worrying is that the exact intensity distribution over the disk/shadow is very algorithm-dependent.''\footnote{Fabien Baron as a judge of Challenge 2, data set 6; see \textcite[162]{bouman_extreme_2017}.} Other data sets resulted in reconstructions with more agreement between imaging methods. In her thesis, Bouman noted:

\begin{quote}
    All algorithms make different imaging assumptions, and thus produce different images, even in the best cases. However, despite these differences, there is often some consensus as to the primary structure in the image. In fact, comparing independently reconstructed images resulting from different methods can often help in identifying which structures are most likely to be real and which are spurious artifacts.\footnote{\textcite[184]{bouman_extreme_2017}.}
\end{quote}

In this way, as researchers were getting to know the image reconstruction methods they were developing, the comparison between different images was seen as helpful for distinguishing between the real aspects and the artifacts.

Having trained on synthetic data during the Imaging Challenges, the Imaging Group now applied its methods to the observational data from the 2017 campaign. The group was divided into four teams working with different algorithms, but using the same data. They had agreed that no team was allowed to look at images made by other teams. However, researchers could submit images to Bouman's website, which would then automatically compare them with one another and produce a score to express their degree of similarity. In this way, researchers would have an idea of how similar the images were without needing to \emph{look} at the images. The members of the Imaging Group hoped that this (among other measures) would prevent bias. If all teams produced similar images, while using different imaging methods and working in isolation from one another, it was considered likely that the images were accurate. The researchers in the Imaging Group called this approach of making images without seeing other images ``blind imaging.'' Images were only made visible to the entire group when its members had become convinced that they had produced similar images with their different approaches. This sharing of images was called the ``non-blind'' stage (Section~\ref{fiducial}).

At first glance, this approach resembles the practices of independent verification in other fields of science. For instance, the Large Hadron Collider (LHC) at CERN has two major overlapping experiments, ATLAS and CMS. To ensure independence, the members of each experiment are not allowed to share certain kinds of information with members of the other, at least in principle, even though this information barrier is not completely impermeable in practice. So, the EHT's ``blind imaging'' approach may be seen as somewhat analogous to the LHC's practice of independent verification. However, it is \emph{not} analogous to what the LHC would call a `blind analysis,' which requires that researchers commit to a specific analysis strategy for a specific section of the data in advance, before `unblinding' and being able to see the data and the results of the analysis. In the EHT's case, the analysis was not `blind' in this sense. Each of the four teams in the Imaging Group had full access to the data, and could freely modify their analysis as many times as they wished. Moreover, each team could even see numerical scores of the degree of similarity between its images and those of the other teams.

\begin{figure}[t]
    \centering
    \includegraphics[width=0.6\textwidth]{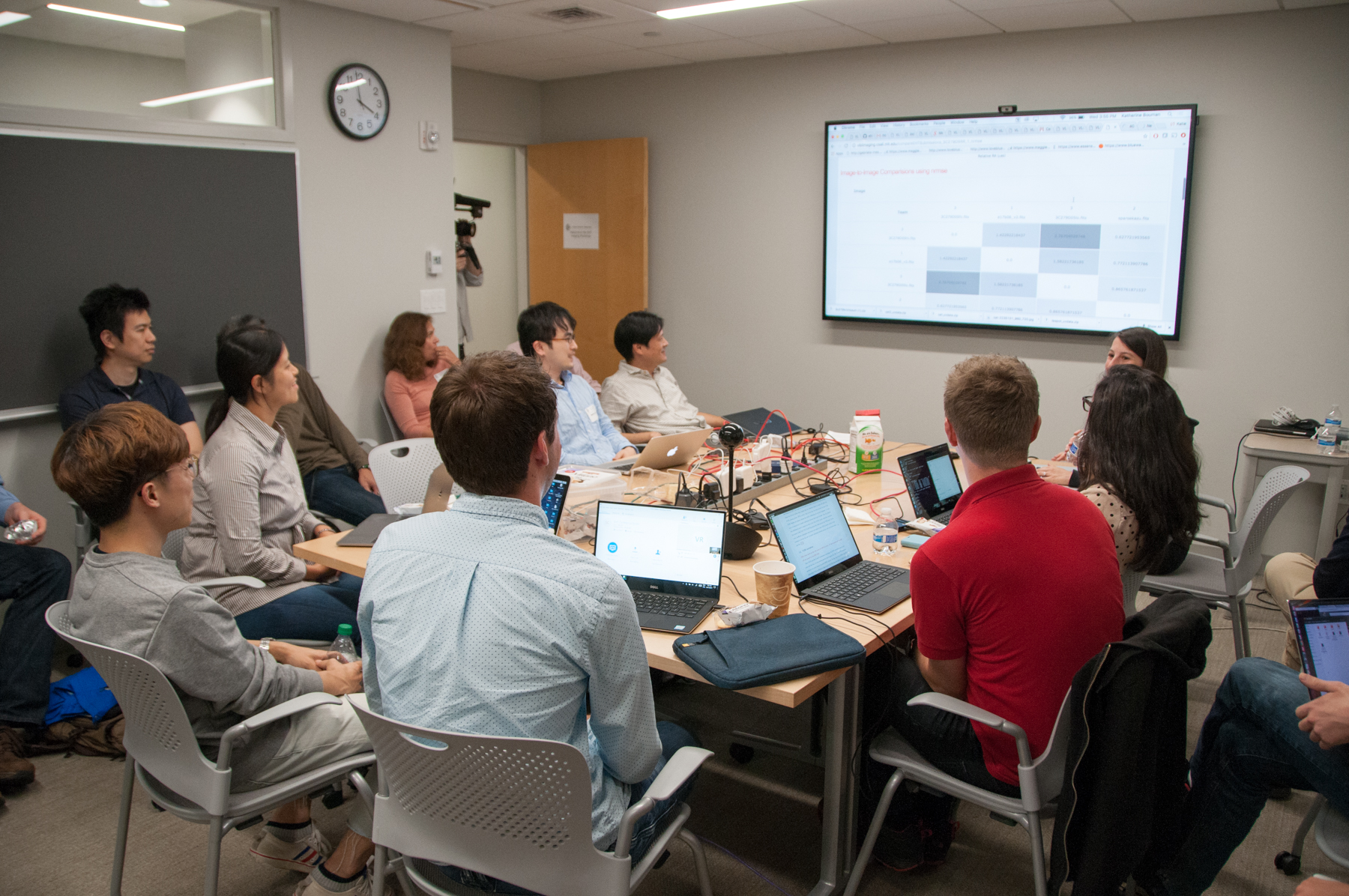}
    \caption{Members of the EHT Imaging Group during a ``blind comparison'' during the first imaging workshop, which took place from October 10 to 13, 2017. Courtesy of Chi-kwan Chan.}
    \label{fig:blind-comparison-photo}
\end{figure}

Phrases capturing the epistemic role of vision, or the lack of it, have a long history in radio astronomy. The field itself, for example, was referred to as ``blind astronomy'' as early as 1949 by physicist Patrick Blackett, in contrast to optical astronomy.\footnote{\textcite[1--2, footnote 8 on 2, and footnote 31 on 425]{sullivan_cosmic_2009}.} `Blind' astronomy could see under different conditions, and reveal objects optical astronomy could not. Of course, the term `blind' also appears in other scientific and scholarly contexts, expressing different meanings. In LIGO's research on gravitational waves, for example, the term ``blind injections'' was used to describe the fake signals that were intentionally introduced in the data as a way to test the collaboration's methods.\footnote{\textcite[93--94]{collins_gravitys_2010}; \textcite[197--200]{collins_gravitys_2013}; \textcite[76--82]{collins_gravitys_2017}.} With the EHT ``Imaging Challenges,'' researchers did not know what they were imaging with their algorithms, and sometimes were deliberately offered data from spurious images, such as those of a snowman instead of a black hole simulation. Both at the training stage during the challenges (Figure~\ref{fig:blind-comparison-photo}), and at the first stage of the actual imaging, the phrase ``blind imaging'' was used; in the latter case, because the four teams that worked in parallel initially remained `blind' to each other's results.

In the LIGO case, we find that use of the term `blind' referred to \emph{concealed knowledge}. In nuclear and particle physics, as we discussed with the LHC, the phrase `blind analysis' is intended to capture \emph{impartiality}; in particular, that measurements are analyzed without bias toward the experimenter's preconceptions or expectations.\footnote{\textcite{klein_blind_2005}.} Clearly, the metaphor of blindness is invoked in different ways in physics, and in scholarship more broadly (e.g., `blind review').\footnote{`Blind review' has been criticized for its use of disability as metaphor, with suggestions of replacing the term with `anonymous review' or `identity-hidden review'; see, e.g., \textcite{ades_end_2020}.} What unites these uses is an association with methodological cross-checking, with the aim of improving reliability and avoiding bias. The EHT's use of ``blindness''---hiding target images during the Imaging Challenges, and each team's results from the others---fell closer to LIGO's sense of concealed knowledge than to the LHC's emphasis on impartiality. 

\section{``Fiducial'' imaging} \label{fiducial}

The members of the Imaging Group were worried that they might reproduce what they expected to see. As we have seen in the previous section, already before imaging from data began, the group tested how well its members could depict what was actually there in the ``ground truth'' images of test data. We also saw how they turned to what they called ``blind imaging'' during the early stage of imaging as part of their aims for objectivity.\footnote{For the use of the terminology of ``ground truth'' and ``blind imaging,'' see, e.g., \textcite[1, 9]{event_horizon_telescope_collaboration_first_2019-4}.} Yet these measures were not sufficient to address all of the researchers' worries. There are countless ways of using the same imaging algorithm. Each algorithm requires researchers to choose values for several parameters, and those parameter choices are not easy to determine. During the ``blind'' stage, researchers were free to manipulate parameters as they wished, exercising their expert judgment and making choices by trial and error. This raised the concern that such choices were too subjective.

After the images were made visible across teams, in the ``non-blind stage,'' the researchers wanted to produce more trustworthy images by choosing the parameters more ``objectively.'' In their words, they sought to ``objectively evaluate the fidelity of the images'' and to ``select imaging parameters that were independent of expert judgment.''\footnote{\textcite[9]{event_horizon_telescope_collaboration_first_2019-4}.} To do so, the researchers returned to synthetic data.

In computational imaging, the selection of optimal parameters typically requires testing their performance on trusted ``ground truth'' data. However, in the case of the EHT, there were no previous trusted images of black holes. In the absence of conventional ``ground truth'' data, the EHT tested its algorithms and parameters on synthetic data. First, the researchers generated a set of synthetic images, including geometric shapes and GRMHD simulations. Then, they produced synthetic visibility data from those images; i.e., they converted the images into visibility data in the same format as EHT observations, as if the telescope array had observed the synthetic images in the sky. This enabled the researchers to test their imaging algorithms and parameters, as if the synthetic data were the ``ground truth.'' Because the original images were known (after all, they were synthetic), the researchers could test the performance of an imaging algorithm and choice of parameters by comparing how closely each reconstruction resembled the original.

\begin{figure}[t]
    \centering
    \includegraphics[width=0.6\textwidth]{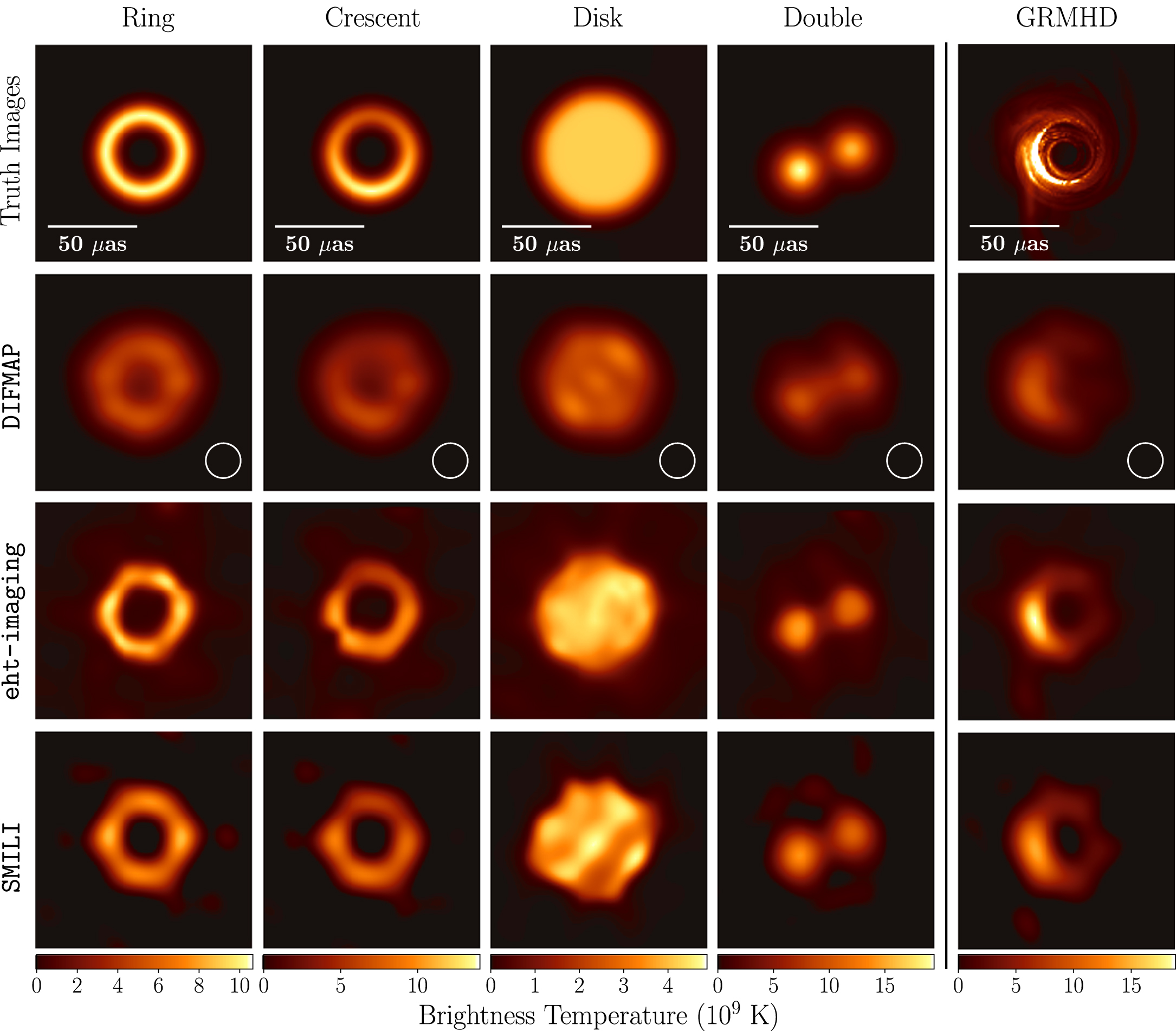}
    \caption{Geometric shapes used as ``truth images,'' alongside reconstructions by different algorithms. Source: \textcite[16]{event_horizon_telescope_collaboration_first_2019-4}.}
    \label{fig:geometric-reconstructions}
\end{figure}

Such tests helped to ensure that a given algorithm and choice of parameters did not produce a ring when the original image did not have a ring. This was important because a ring, with asymmetry and brightness depression (shadow), had been predicted and was what researchers likely expected after having looked at a multitude of visualizations from simulations of the immediate surroundings of black holes. Because of this, the geometric shapes used as ground truths included not only rings but also crescents, filled disks, and double point sources of which one source was slightly brighter (Figure~\ref{fig:geometric-reconstructions}).

\begin{figure}[t]
    \centering
    \includegraphics[width=\textwidth]{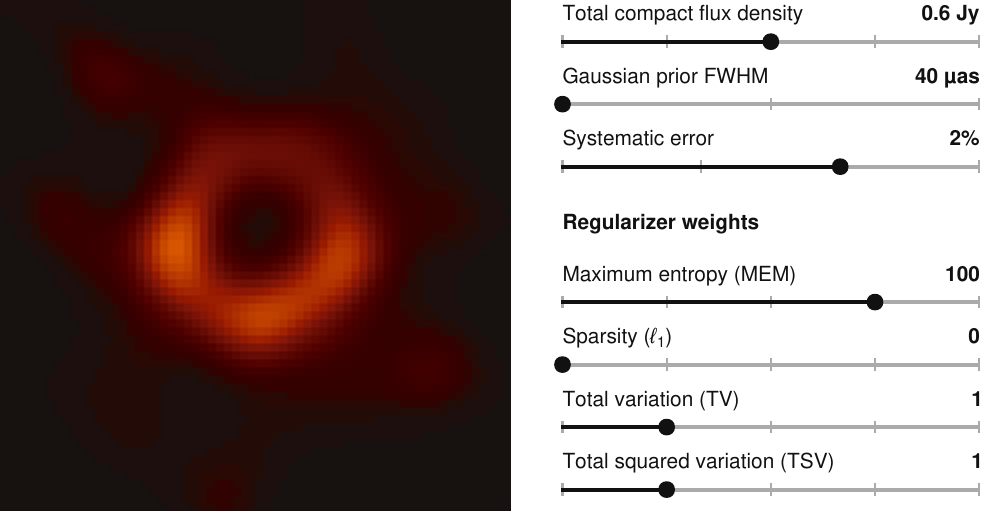}
    \caption{Preview of an interactive demonstration of parameter choices for an RML algorithm, based on the software package \emph{eht-imaging}. Following the EHT's ``parameter survey,'' it samples seven parameters, totaling 37,500 possible combinations. We computed the images for all combinations in advance; the demonstration displays these precomputed results rather than running the algorithm live. The interactive version, in which the reader can vary the parameters and compare the resulting images, is available at \url{https://nbarchive.ku.dk/resources/coloring-black-holes/} and archived at \url{https://doi.org/10.5281/zenodo.21328092}.}
    \label{fig:interactive-rml}
\end{figure}

The Imaging Group conducted a ``parameter survey'': it automatically tested many possible combinations of parameters for a given algorithm in order to pick the combination that performed best (see the interactive version of Figure~\ref{fig:interactive-rml} to vary these parameters and compare the resulting images). The researchers had to decide which parameters to include in the survey. In particular, for the RML algorithms, they had to decide which regularizers to include. One could argue that an epistemically conservative approach would seek to minimize the number of regularizers, since regularization increases the risk of imaging artifacts. However, the survey took a maximalist approach, including many kinds of regularizers.

This maximalist approach also made the survey more computationally expensive. A single run of an imaging algorithm with one parameter combination is already costly, and the number of combinations increases exponentially with the number of parameters. Because of the computational cost, the researchers were not able to make a fine-grained survey. To establish the weight of regularizers, the survey tested only by orders of magnitude, varying the weight of each parameter in powers of 10. Each of the three imaging pipelines had different parameters and was surveyed separately. To illustrate: for one of the algorithms (implemented by a software package called \emph{eht-imaging}), the survey tested five values for the weight of each regularizer (0, 1, 10, 100, 1000), finding that the best-performing combination weighted one regularizer (maximum entropy, based on a circular Gaussian prior) 100 times more heavily than the others. Since the three pipelines used different regularizers and parameter structures, these weights did not transfer across pipelines, and the coarse-grained nature of the survey---constrained by computational cost---left open whether finer-grained tuning would have changed the results.

Although the researchers ultimately selected a single parameter combination per pipeline to produce each fiducial image, the survey also identified a broader ``Top Set'' of well-performing combinations---for \emph{eht-imaging}, this included 1,572 of the 37,500 tested combinations.\footnote{\textcite[17]{event_horizon_telescope_collaboration_first_2019-4}.} The stability of image features (particularly a ring) across the Top Set provided evidence that the results were not overly sensitive to the precise choice of parameters. The researchers called the single best-performing combination for each pipeline ``fiducial parameters'' and the resulting images ``fiducial images,'' and argued that the surveys on synthetic data allowed them to ``objectively evaluate the fidelity of the images.''\footnote{\textcite[2, 9]{event_horizon_telescope_collaboration_first_2019-4}.}

They did not explain the reasoning behind this terminological choice, but the word `fiducial' has a long history in astronomy. In addition to the broader meaning ``of or pertaining to, or of the nature of, trust or reliance'' (common, for example, in theology), the Oxford English Dictionary lists a more specific meaning dating back to the 16th century: ``In \emph{Surveying}, \emph{Astronomy}, etc. Of a line, point, etc.: Assumed as a fixed basis of comparison.''\footnote{See \textcite{noauthor_fiducial_2025}.} This meaning is also common in photography, microscopy, and computational imaging: a `fiducial marker' is an object that is used as a point of reference, for example a ruler or a barcode tag placed in the field of view of a camera.\footnote{See, e.g., \textcite{fiala_artag_2005}; \textcite{mohammadian_high_2019}.} So, the EHT's use of ``fiducial'' seems somewhat unusual, because it did not refer to the resulting images or parameters that constituted the fixed basis for comparison. Rather, its synthetic data played the `fiducial' role, as `fiducial marker.' The EHT called the resulting images themselves ``fiducial.''

After finding the best-performing parameters based on synthetic data, the Imaging Group applied these parameters to the observational data of M87* to generate new images. The Imaging Group generated three ``fiducial images'' of M87*, resulting from three imaging pipelines. One of the pipelines used a CLEAN-based algorithm as implemented by Difmap. The other two pipelines used RML algorithms, but implemented in slightly different ways by two software packages, both of which were developed specifically for the EHT. The software package \emph{eht-imaging} (or \emph{ehtim}) was developed by many and managed primarily by Andrew Chael, and the package SMILI had four core developers---Kazu Akiyama, Fumie Tazaki, Shiro Ikeda, and Kotaro Moriyama---at the time of the M87* publication.\footnote{\textcite{akiyama_smili_2019}; \textcite{chael_simulating_2019}; \textcite{chael_ehtim_2019}.}

\begin{figure}[t]
    \centering
    \includegraphics[width=0.6\textwidth]{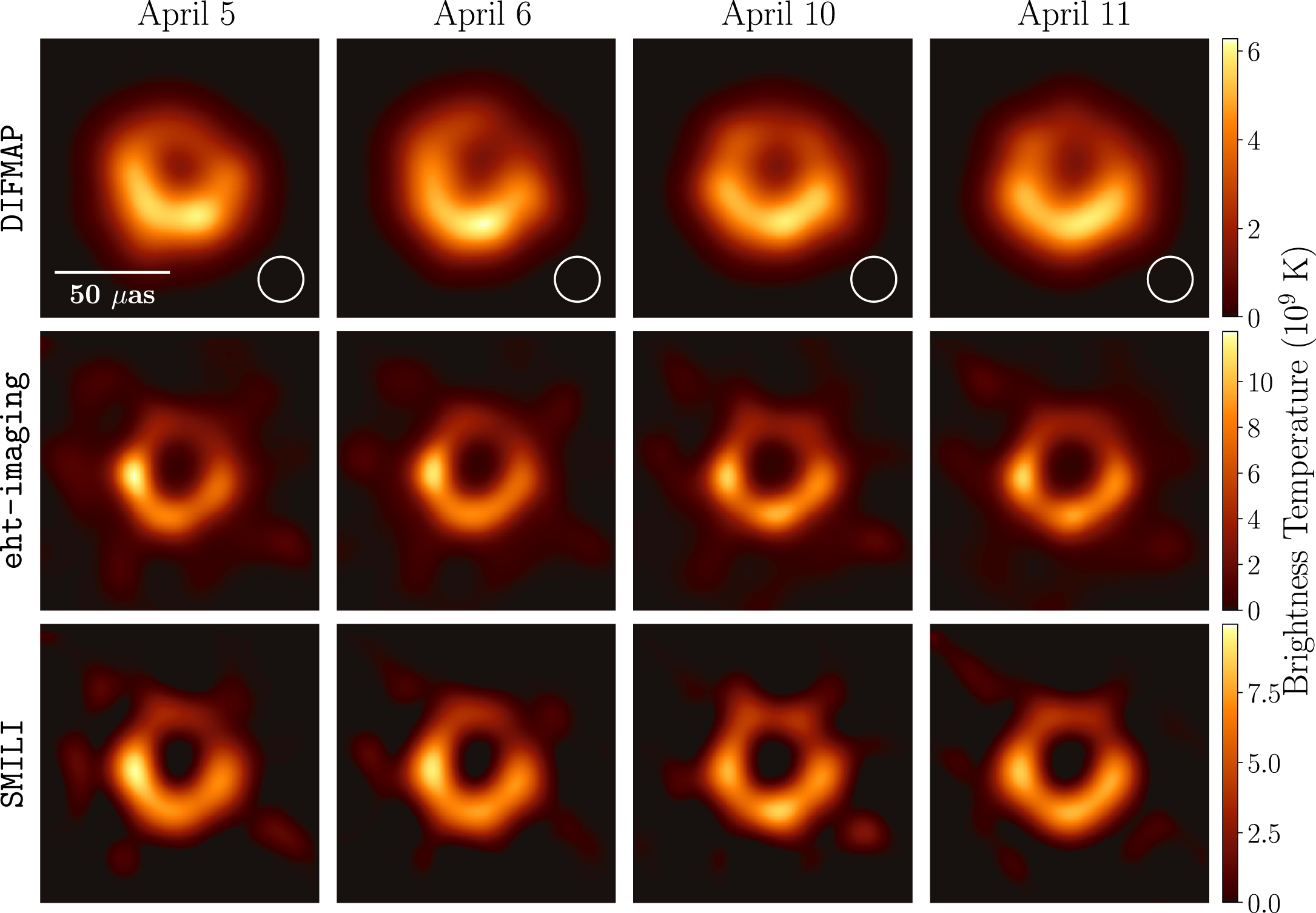}
    \caption{Reconstructions of M87* by three different algorithms for each of the observing days in April 2017. Source: \textcite[21]{event_horizon_telescope_collaboration_first_2019-4}.}
    \label{fig:m87-comparison}
\end{figure}

Figure~\ref{fig:m87-comparison} shows images reproduced by the three algorithms for each of the observing days in April 2017. In describing these results, the authors of the imaging paper (Paper IV) noted that the images were ``broadly consistent'' in their asymmetric ring and brightness depression.\footnote{\textcite[18]{event_horizon_telescope_collaboration_first_2019-4}.} The methods did not, however, reproduce identical images, as was visually displayed in Figure~\ref{fig:m87-comparison} and emphasized textually: ``[a]s expected from tests on synthetic data [...] the details of the reconstructions differ between the imaging methods.''\footnote{\textcite[19]{event_horizon_telescope_collaboration_first_2019-4}.}

The EHT's imaging process involved synthetic data at multiple stages, raising questions about the relationship between observation and simulation. 
On the one hand, the resulting images are not mere simulations in the sense that GRMHD simulations are. After the parameters are selected on the basis of synthetic data, the parameters are applied to the observational data of M87*. The researchers also tested the parameters on other observational data from more trusted sources, particularly the quasar 3C 279. On the other hand, the resulting images are not straightforward observations either. Without the complex interplay of observational and simulated data at many stages, the resulting images would not be possible---though, as we will see, not all uses of simulations are equally consequential.\footnote{Historians and philosophers of science have long discussed the role of simulation, including its relation to theory and experiment. See, e.g., \textcite{galison_image_1997}; \textcite{humphreys_extending_2004}; \textcite{morgan_experiments_2005}; \textcite{radder_philosophy_2003}; \textcite{rohrlich_computer_1990}. \textcite[601]{frigg_philosophy_2009} argue that simulations do not require a wholly new epistemology, but that ``models are more complex than traditional philosophy of science allows'' and that ``we still do not have a worked out epistemology that accounts for this.'' The entanglement of observational and simulated data is not unique to the EHT; for a case involving the use of models to correct observational data in paleontology, see \textcite{bokulich_using_2021}. For a recent analysis of theory-ladenness in black hole observations, see \textcite{doboszewski_theory_2025}.}

\begin{figure}[t]
    \centering
    \includegraphics[width=\textwidth]{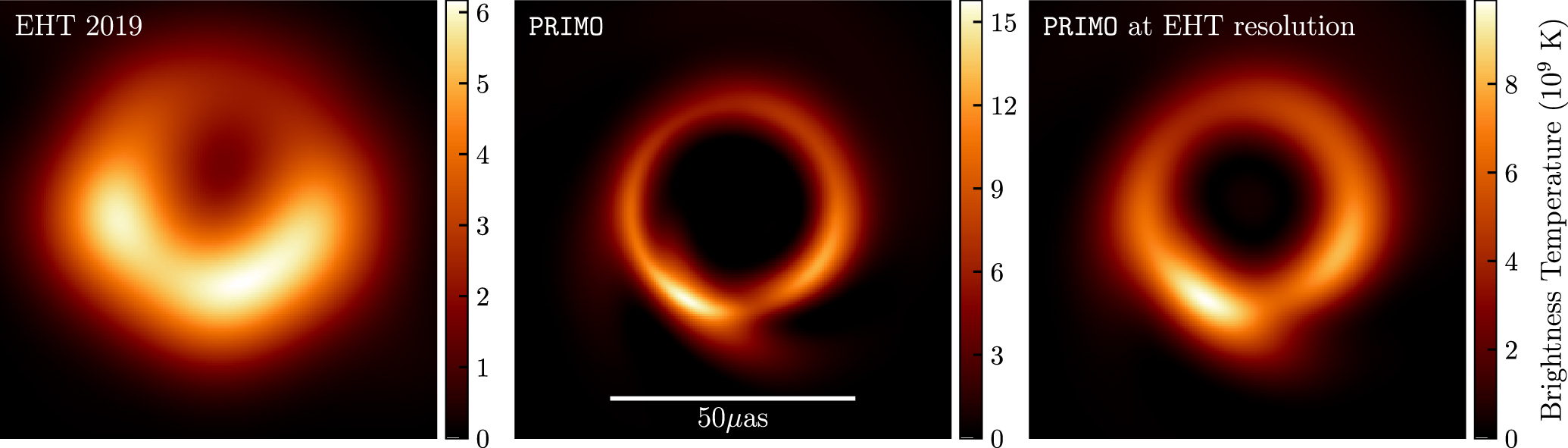}
    \caption{New reconstructions of M87* made using the machine-learning-based PRIMO algorithm in 2023 (center and right), compared with the EHT's image from 2019 (left). Source: \textcite[2]{medeiros_image_2023}.}
    \label{fig:m87-primo}
\end{figure}

In April 2023, four years after the original M87* publication, four EHT researchers published a new image based on the same data (Figure~\ref{fig:m87-primo}), adding a new twist to this interplay between observation and simulation.\footnote{\textcite{medeiros_image_2023}.} The authors---astrophysicists Lia Medeiros, Dimitrios Psaltis, Tod R. Lauer, and Feryal Özel---developed an algorithm, named PRIMO, based on machine-learning techniques.\footnote{\textcite{medeiros_principal-component_2023}. See also \textcite{sun_deep_2021}.} A machine-learning model was first trained to detect patterns in data from GRMHD simulations, then used these patterns to fill in the gaps in the observational data. According to the authors, this algorithm produced an image with an ``improved resolution.''\footnote{\textcite[3]{medeiros_image_2023}.}

Like the parameter survey and the earlier benchmarking of CLEAN and RML algorithms (discussed in Section~\ref{algorithm}), PRIMO relied on GRMHD simulations. But the earlier uses did not depend on the accuracy of those simulations, which served as test inputs alongside theory-independent geometric shapes. PRIMO's use was different: the machine-learning model reproduced patterns from the simulated images. If those simulations were wrong, PRIMO's output would be compromised in a way that the original EHT images would not. The resulting image raised concerns of circularity: How can the image offer evidence of black holes if it was generated by a model trained on simulations that already assume black holes?

The PRIMO authors sought to address such worries in various ways. They presented their algorithm as complementary to, and motivated by, the others: ``The robustness of the ring-like shapes of the images generated with model-agnostic methods motivates the use of [PRIMO].''\footnote{\textcite[1]{medeiros_image_2023}.} The authors also highlighted the benefits of their machine-learning approach over the others, writing that PRIMO uses ``physically motivated training (as opposed to ad hoc regularizers) to infer the maximum information from the data.''\footnote{\textcite[3--4]{medeiros_image_2023}.} By referring to ``ad hoc regularizers,'' the authors emphasized that ``model-agnostic''\footnote{\textcite[1]{medeiros_image_2023}.} methods like RML also depend on arguably heavy assumptions, not from theoretical physics but about the structure of the image.

In any case, not all EHT researchers were confident in images that rely on this use of machine learning, even if none has published a direct objection. Although the PRIMO authors are EHT members, the collaboration as a whole did not sign the scientific paper by listing its name as an author, and the collaboration did not announce the work through a press release. Instead, the institutions of the individual researchers---the Institute for Advanced Study, the NSF NOIRLab (i.e., the US National Science Foundation National Optical-Infrared Astronomy Research Laboratory), and the Georgia Institute of Technology---issued a joint press release, titled ``A Sharper Look at the First Image of a Black Hole.''\footnote{\textcite{sandberg_sharper_2023}. See also press reports, e.g., \textcite{overbye_that_2023}.} Media reports generally reproduced that release's framing of a ``sharper'' image at ``maximum resolution,'' without mentioning any methodological discussion about the role of simulation.

This is not the only case of machine-learning models in physics research that were first trained on simulations, and then applied to empirical data. Such models have become increasingly common in particle physics, where the simulations are based on theories that have been successful in passing experimental tests and providing precise predictions---the Standard Model---and on additional models tuned to experimental measurements.\footnote{For a survey of uses of machine learning in high-energy physics, see \textcite{calafiura_artificial_2022}.} In the case of the EHT, however, GRMHD simulations cannot be tested against empirical data at the necessary resolution, and it is not a given that all researchers would have the same confidence in detailed images generated by machine-learning models trained on them.

\section{Averaging, coloring, and publication} \label{averaging}

The EHT collaboration decided to publish a single image as its central scientific result. Yet, the EHT did not initially publish the full original data, which was on the scale of petabytes: when the EHT presented the M87* image in its journal publications in April 2019, it also released only a small data set of a few megabytes. This was already calibrated and intended to allow for the reproduction of the same image. Larger processed data sets, on the scale of gigabytes or terabytes, were released only years later and with considerably less fanfare than the collaboration's first image.

The publicized image was not produced by a single algorithm. Instead, it was the simple average of three images, which were produced by the three different imaging pipelines using different algorithms and parameters. The published image of M87* was simply the pixel-by-pixel average of the results from the three pipelines.

Why did the EHT take this unusual approach of averaging three images instead of choosing one---or presenting and emphasizing all three? In his account of the internal discussions, historian of science and EHT collaborator Peter Galison identified three motivations. One was to ``be as conservative as possible'': to minimize the quirks of particular algorithms. Another was ``to test [the averaged image's] relation to other images'': to confirm that the averaged image is qualitatively similar to each of the three constituent images. The third reason was more political in nature rather than intrinsically epistemically motivated: one wished to ``be inclusive of the whole collaboration.'' The EHT was a large international collaboration and ``nobody wanted to be left out'' of the final result, which might result if a single pipeline image had been picked.\footnote{\textcite{galison_philosophy_2019}.}

\begin{figure}[t]
    \centering
    \includegraphics[width=\textwidth]{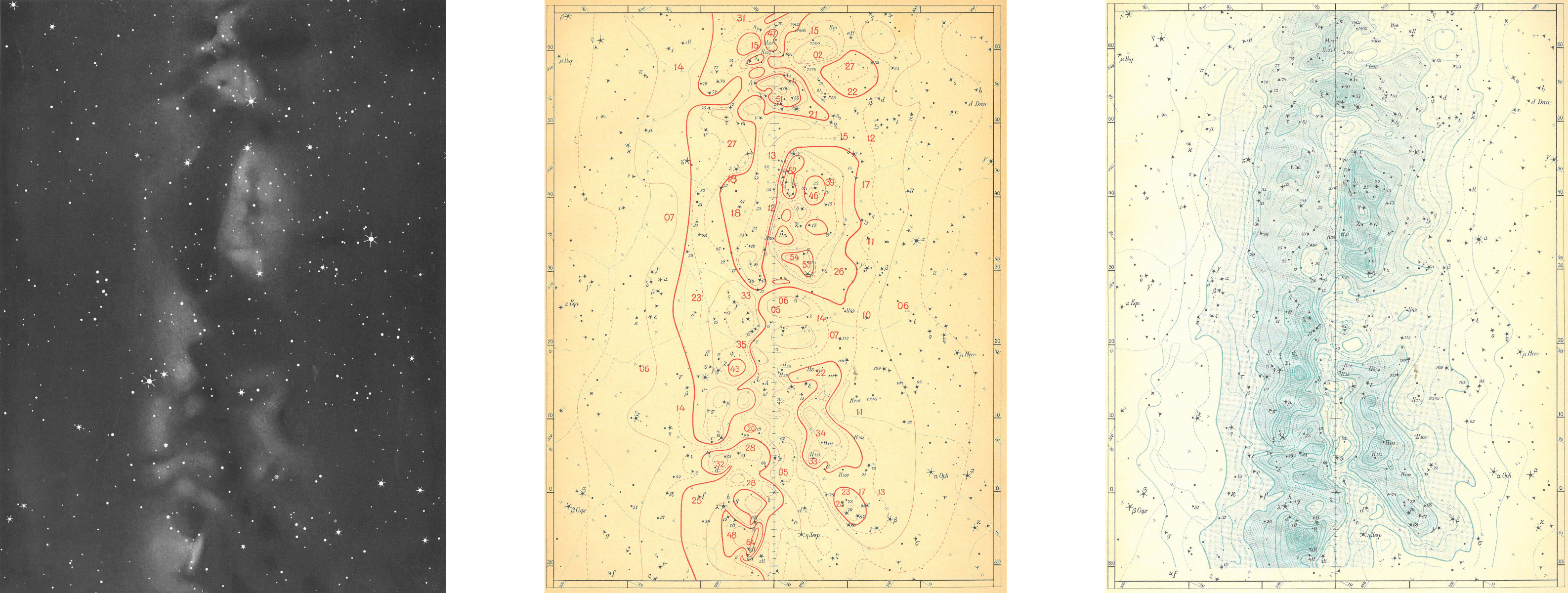}
    \caption{Anton Pannekoek's ``mean subjective image'': as a photograph of the Milky Way offered too much resolution, which loses vital information, Pannekoek made a naturalistic drawing of a region of the northern Milky Way (left), of which he then made an isophotic map (center). After doing the same for drawings made by other observers, he then averaged over these isophotic maps, to finally produce the ``mean subjective image'' (right). Source: \textcite[plates 1, 4, and 7]{pannekoek_nordliche_1920}. See also \textcite[44--48]{tai_anton_2021}.}
    \label{fig:pannekoek-milky-way}
\end{figure}

While unusual, the averaging of images has precedents in the history of astronomy. For example, a century earlier, in 1920, astronomer Anton Pannekoek produced what he called ``the mean subjective image of the Milky Way.'' He set out to \emph{draw} the Milky Way, using human faculties as a kind of recording device to capture the particular idiosyncratic aspects of its appearance. Yet, such drawings would be intrinsically subjective, which Pannekoek considered problematic: to suppress bias, he used a number of observers, who should not be allowed to repeat their drawing of the same portion of the night sky, at least not knowingly; then their work should be averaged over to construct ``the mean subjective image'' (see Figure~\ref{fig:pannekoek-milky-way}). Thus, averaging over images, as in the case of the EHT, was done by Pannekoek to limit the bias introduced by human choices.\footnote{On Pannekoek, see \textcite{tai_left_2017}; \textcite{tai_anton_2016}; \textcite{tai_anton_2019}; \textcite{tai_anton_2021}; \textcite{nasim_labour_2019}.}

Some astronomical imaging choices, however, do not need to be constrained. For example, the EHT's color map of the published image was designed specifically for it, and its choice was only one among many options. Radio waves are invisible to the human eye, and the EHT images were based on mapping different values of estimated brightness temperature to different colors. The EHT ultimately decided on a custom derivative of an existing color map available in the data visualization software package used by EHT scientists (\emph{matplotlib}). The existing version, called \emph{afmhot}, mapped brightness temperature values to colors ranging from black (lowest value) to red to orange to yellow to white (highest). There were multiple reasons behind this choice. First, color maps similar to \emph{afmhot} had been used by some earlier papers with black hole images based entirely on simulation, for example in the 2000 paper by Heino Falcke, Fulvio Melia, and Eric Agol (mentioned in Section~\ref{setup}). Thus, scientists interested in black hole imaging were already accustomed to seeing and interpreting similar colors. Even though the EHT reached a final decision on color use only in the last months before publication, many EHT researchers had already been using \emph{afmhot} and similar color maps at earlier stages of the project.

Yet, the choice was also motivated by common associations in visual culture. Because the colors of \emph{afmhot} are commonly associated with warmth or hotness, EHT scientists found them appropriate to convey the high temperature of a black hole's surrounding emission region. Both scientists and the press often invoked the metaphor of a ``ring of fire'' to describe radiation emitted by the plasma surrounding the black hole shadow.\footnote{\textcite{castelvecchi_black_2019}; \textcite{clery_for_2019}; \textcite{ghosh_first_2019}. For a different case of how color choices in astronomical imaging are justified epistemically while drawing on broader visual culture, see \textcite{kessler_technologys_2021} on the ``hyperchromatic'' or ``psychedelic'' color palette of Jupiter and Saturn images by NASA's Voyager mission.} Falcke did so as he announced the image at the official European press conference, when he also commented that ``it feels like really looking at the gates of Hell, at the end of space and time.''\footnote{\textcite[timestamps: 0:33 for ``ring of fire'' and 2:18 for ``gates of Hell'']{event_horizon_telescope_collaboration_breakthrough_2019}.} It should be noted that in addition to his scientific work, Falcke is an ordained lay minister in the Protestant Church in Germany.\footnote{Falcke's choice of words has been deliberate, as he explains in a recent interview (in Dutch, our translation): ``On the one hand these kinds of references are tongue-in-cheek, on the other hand I also make them to point out the relation between the natural sciences and the way of thinking in Judeo-Christian thought,'' in particular as both imply a reliable and predictable (as opposed to magical) universe, rooted in a shared heritage; see \textcite[28]{keulen_2025}.}

Many other options, including custom color maps made by EHT researchers, were also available. Here are a few alternatives (Figure~\ref{fig:m87-recolored}):

\begin{figure}[t]
    \centering
    \includegraphics[width=0.6\textwidth]{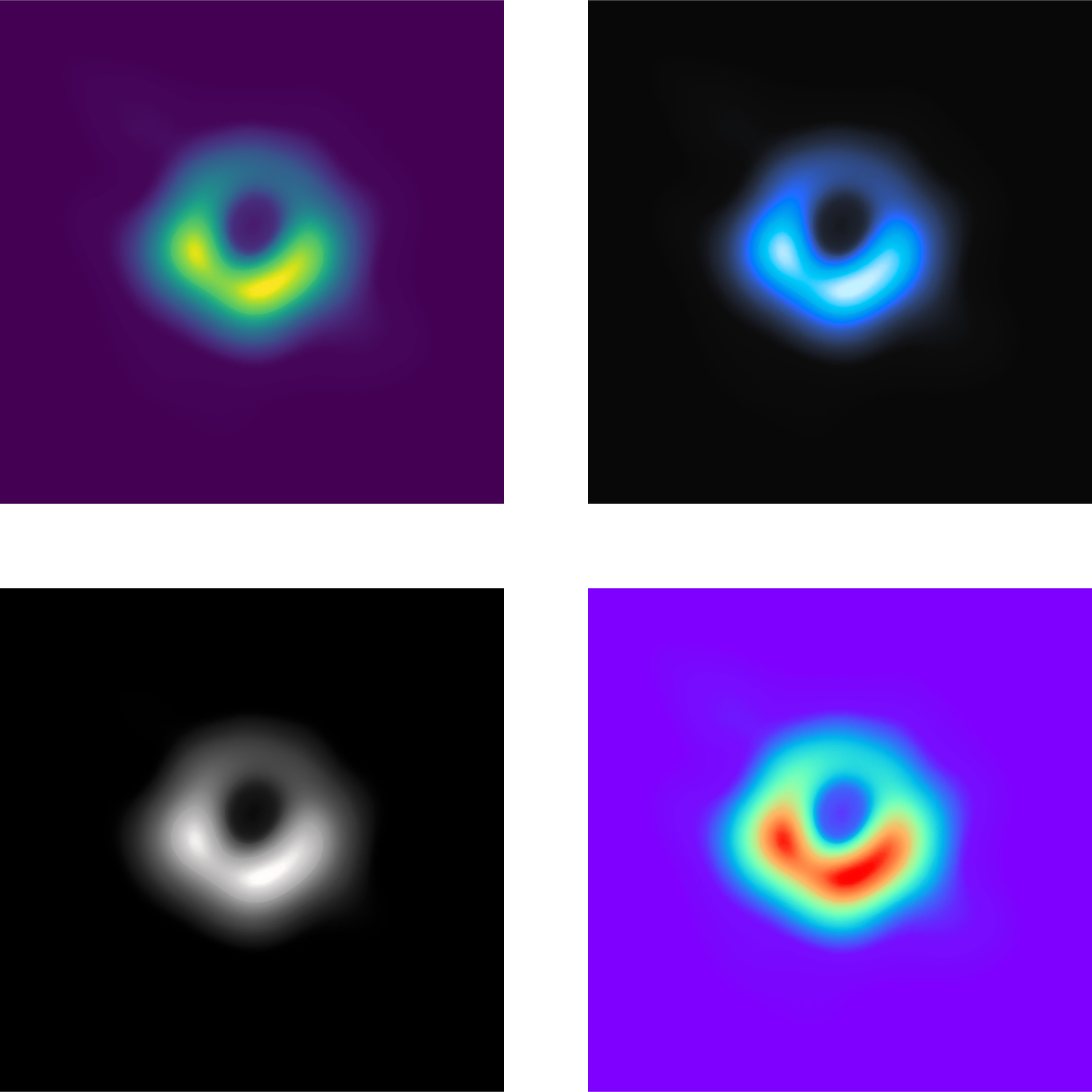}
    \caption{The M87* image in alternative color maps. These images were generated by recoloring the final published image using color maps from the EHT's software. The blue (top right) and grayscale (bottom left) color maps were among the 196 options that were custom-made for the EHT by a group led by Chi-kwan Chan and Lia Medeiros. The ``viridis'' map (top left) is the default option in the \emph{matplotlib} software package used by the EHT, and the ``rainbow'' map (bottom right) is also included in the package.}
    \label{fig:m87-recolored}
\end{figure}

\begin{figure}[t]
    \centering
    \includegraphics[width=0.8\textwidth]{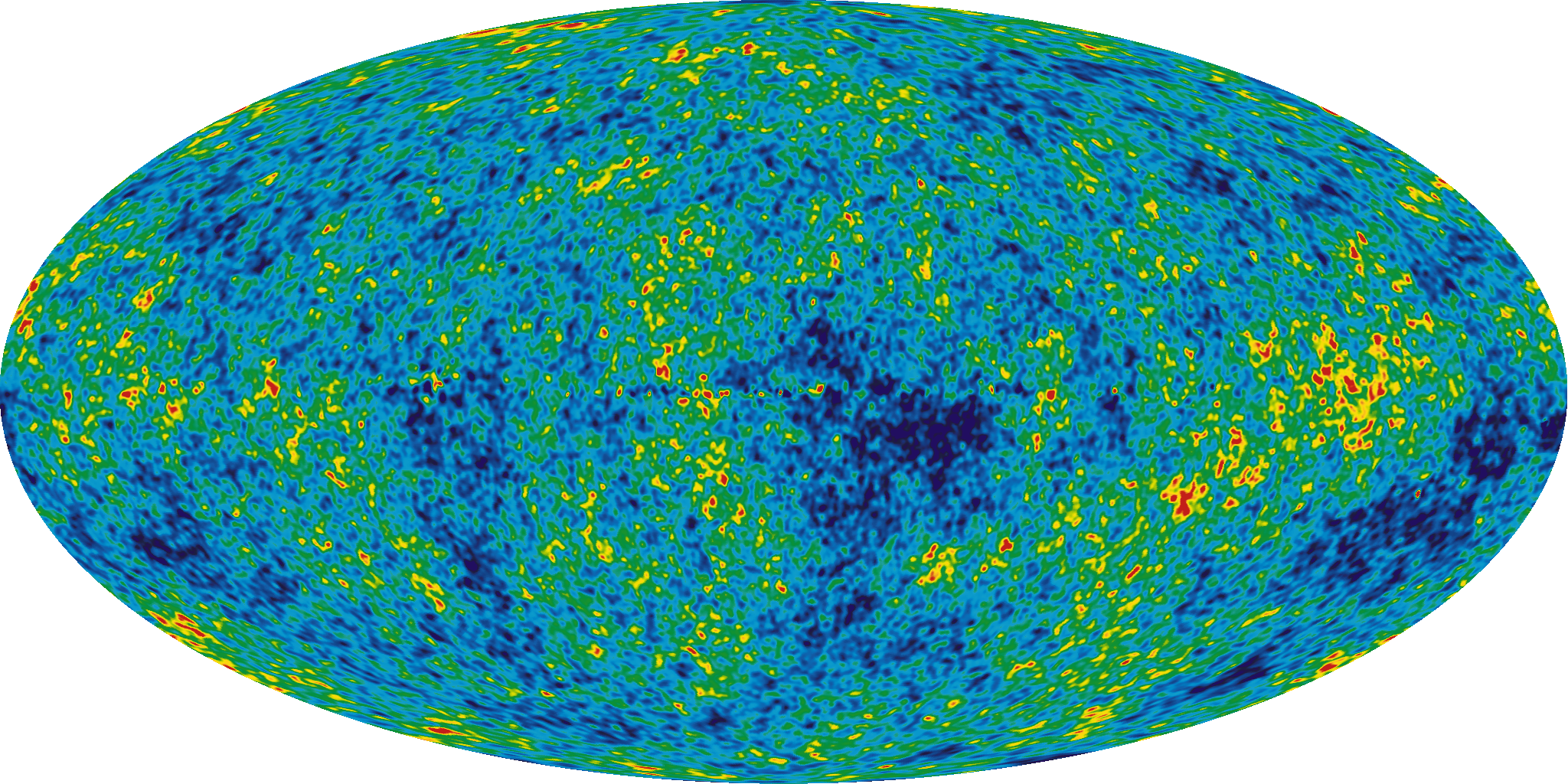}
    \caption{Heat map of temperature fluctuations in the cosmic microwave background, based on nine years of Wilkinson Microwave Anisotropy Probe data (2012). It uses the rainbow color map. Credit: NASA.}
    \label{fig:cmb-map}
\end{figure}

\begin{itemize}
\item
  \emph{Default.} A color map called ``viridis'' was the default choice in the \emph{matplotlib} software package. We do not know of any researchers who argued for using it. But if the code had not specified any color map at all, the software would have rendered the data in this one.
\item
  \emph{Blue.} Some EHT researchers argued that blue would be a more appropriate choice to convey hotness, since the hottest flame is blue rather than orange.
\item
  \emph{Grayscale}. Because the values were limited to one dimension (brightness temperature), grayscale would be a reasonable choice because it would intuitively affirm the notion that one is dealing with only one dimension.
\item
  \emph{Rainbow.} The rainbow color map varies hue instead of brightness or luminosity. It is widely used for scientific visualization in many fields, from oceanography to neurobiology.\footnote{\textcite{helmreich_colors_2021}; \textcite[93]{dumit_picturing_2004}.} In cosmology, the rainbow color map and similar hue-varying maps have been used in images of the cosmic microwave background (see Figure~\ref{fig:cmb-map}).\footnote{\textcite{doran_instrumentalizing_2021}.}
\end{itemize}

According to the EHT's open-source code repositories, a group of EHT researchers led by Chi-kwan Chan and Lia Medeiros made 196 custom color maps, even though there were already some 80 color maps built into the \emph{matplotlib} package.\footnote{The color maps are included in the software package \emph{ehtplot}. The code is available in \textcite{chan_ehtplot_2021}.} Some custom maps were derivatives of built-in maps; others were entirely new. EHT researchers sought to comply with standards informed by the study of human visual perception. Some such standards have long histories, dating back to the emergence of color science around the end of the 19th century.\footnote{On the history of American color science, see \textcite{rossi_republic_2019}.} Although many standards exist, the EHT focused on two in particular. The first was one of the oldest and arguably the most commonly adopted: `perceptual uniformity.' A perceptually uniform color map is designed so that equal differences in value will be perceived as equal differences in color to a typical human subject. Most color maps are not perceptually uniform unless intentionally designed as such. For example, the most conventional version of the rainbow color map is not perceptually uniform because humans tend to perceive differences between shades of green, for instance, as much smaller than between shades of blue; such claims of uniformity are informed by findings from human experiments conducted by psychologists. The other standard that the EHT collaboration aimed for, and which is less commonly adopted, is `symmetry in chroma.' This meant that all colors in the map should have the same degree of purity or intensity, i.e., no color should appear more vivid or dull than another. The published image used a custom variant of the \emph{afmhot} color map that satisfied these two standards, `perceptual uniformity' and `symmetry in chroma.'

In preparing the image for publication, the EHT made several deliberate choices so that the image would resemble a conventional photograph more closely. Among the hundreds of available color maps, most of which would not have produced such a resemblance, the EHT picked one that did so quite effectively. The chosen color map has a black background (most other maps do not) resembling the night sky, and the red/orange/yellow gradient allowed the black hole shadow to be described evocatively as a ``ring of fire.'' Moreover, the publicly released final image had a landscape aspect ratio, which is the most common in photography, even though the images produced during the earlier stages had typically been displayed in a square format. Since the EHT's software produces square images when using the code and data provided with the papers, the outputs had to be modified for presentation in a landscape format.\footnote{On the landscape tradition and astronomical imaging, see \textcite{kessler_pretty_2011}; \textcite{kessler_picturing_2012}; \textcite{messeri_placing_2016}; \textcite{vertesi_seeing_2015}.}

When Figure~\ref{fig:m87-public} was released, it was often referred to in communications targeting the general public as a ``picture'' or ``photo,'' as when Doeleman stated that the EHT had ``taken a picture of a black hole.''\footnote{\textcite[timestamp: 7:38]{event_horizon_telescope_collaboration_and_national_science_foundation_nsf_2019}. Phrasings such as ``first photo of a black hole'' were widely used by science writers and journalists following the EHT's release. See, e.g., \textcite{akpan_here_2019}; \textcite{strickland_fuzzy_2023}.} The notion of ``picture'' is intuitively understood by the public as an image obtained by a quick and straightforward process. Calling the image a ``picture,'' then, could be understood as part of a rhetorical strategy aimed at convincing audiences of the accuracy of the depiction: it ties the image to a familiar and supposedly ``direct'' way of capturing more mundane visual phenomena.\footnote{On the notion of ``directness'' in black hole imaging, see \textcite{skulberg_what_2025}. Directness was also central in the communication of gravitational wave research. See \textcite[93--94]{collins_gravitys_2010}; \textcite[197--200]{collins_gravitys_2013}; \textcite[76--82]{collins_gravitys_2017}; \textcite{elder_direct_2025}; \textcite{franklin_is_2017}. On the history and philosophy of gravitational wave research, see also \textcite{kennefick_traveling_2007}; \textcite{blum_gravitational_2018}; \textcite{bonolis_gravitational-wave_2020}; \textcite{patton_expanding_2020}.} Doeleman further emphasized the visual nature of the evidence:

\begin{quote}
    What you are seeing is evidence of an event horizon. By laying a ruler across this black hole, we now have visual evidence for a black hole. We now know that a black hole that weighs 6.5 billion times what our Sun does exists in the center of M87, and this is the strongest evidence that we have to date for the existence of black holes.\footnote{\textcite[timestamp: 8:09]{event_horizon_telescope_collaboration_and_national_science_foundation_nsf_2019}.}
\end{quote}

This reflects a long history of viewing images as epistemically superior evidence, both in astronomy and more broadly.\footnote{\textcite[15, 351]{sullivan_cosmic_2009}; \textcite{galison_artificial_1988}; \textcite[65--73]{galison_image_1997}.} In the peer-reviewed papers, by contrast, the words ``photograph'' and ``picture'' were not used; the figures were instead referred to as ``images,'' and the carefulness of the process was emphasized to speak for their accuracy.\footnote{\textcite{event_horizon_telescope_collaboration_first_2019-1}; \textcite{event_horizon_telescope_collaboration_first_2019-2}; \textcite{event_horizon_telescope_collaboration_first_2019-3}; \textcite{event_horizon_telescope_collaboration_first_2019-4}; \textcite{event_horizon_telescope_collaboration_first_2019-5}; \textcite{event_horizon_telescope_collaboration_first_2019-6}.}

Visual and textual rhetoric is a part of the promotion of all large-scale science projects and their results, from the LIGO-Virgo Collaboration emphasizing directness, to Fermilab focusing on the notion of the frontier, to some of the Hubble Telescope's images invoking Romantic imagery of the American West.\footnote{On the rhetoric of big science, see \textcite{martin_critiques_2023}; \textcite{hoddeson_fermilab_2008}. On invoking cultural traditions in connection with the Hubble Space Telescope, see \textcite{kessler_picturing_2012}. On big science, see also \textcite{baneke_david_lets_2020}; \textcite{baneke_big_2023}; \textcite{capshew_big_1992}; \textcite{galison_big_1992}; \textcite{kragh_origins_2023}; \textcite{kragh_big_2024}; \textcite{mccray_giant_2004}; \textcite{hoddeson_fermilab_2008}; \textcite{riordan_tunnel_2015}; \textcite{smith_space_1989}; \textcite{westfall_rethinking_2003}.} It is also common for large-scale projects in astronomy to communicate to different audiences in different ways.\footnote{In fact, in some ways the EHT chose to aim for a shared visuality when communicating to different audiences, for example by using the same color map. See \textcite{skulberg_event_2021}.} We have shown how the EHT's choice to announce a single image as its final result, which was key to its rhetorical approach, oriented the scientific work at several stages toward convergence on such an image. At the final stage, researchers had to average three different images produced with different algorithms, and to choose a single color map among hundreds of possibilities.

\section{Conclusion} \label{conclusion}

How, then, should we interpret black hole images? Although we emphasize that the framing of the published image as a ``photo'' has played an important rhetorical role in the EHT's communication strategy, we do not mean to argue that the image is \emph{not} a photograph. This would put us on the path of a merely definitional dispute, while, as Galison has put it, ``it seems that chasing after a fixed definition of photography, with a once-and-forever set of criteria, is a losing proposition.''\footnote{\textcite{galison_how_2021}.} Indeed, the category of photography has been sufficiently vibrant and capacious to encompass diverse kinds of images, some of which have been made without a camera, in media other than film, at infrared wavelengths, or based on silhouettes. The making of a digital photograph can also involve various human choices (think of focal length, aperture, and other settings on a camera) as well as complex algorithms for data processing. Yet, we note that there is an important difference: in the case of the EHT, the algorithms could not be validated by testing their outputs against previously trusted images at the same extreme limits of resolution, since there were none. 

Similarly, although the EHT images blend empirical and synthetic data in ways that blur the boundary between observation and simulation, we do not mean to argue that the results should not count as observations. As historians of science Lorraine Daston and Elizabeth Lunbeck write, scientific observation has ``a long, surprising, and epistemologically significant history, full of innovations that have enlarged the possibilities of perception, judgment, and reason''; among those innovations are instruments that ``include not only the naked senses, but also tools such as the telescope and microscope, the questionnaire, the photographic plate, the glassed-in beehive, the Geiger counter, and a myriad of other ingenious inventions designed to make the invisible visible, the evanescent permanent, the abstract concrete.''\footnote{\textcite[1--2]{daston_introduction_2011}.} Indeed, black hole imaging is part of an ongoing episode in this history, one in which scientists in many fields are enlarging the category of `observation' to include forms of evidence which rely ever more heavily on simulation techniques and synthetic data.

Instead of disputing the definitions of `observation' or `photograph,' we have analyzed the assumptions and choices involved at each stage of data processing.
Philosopher Nora Mills Boyd has noted that processing in radio astronomy is a pertinent example of how ``[t]he target system under study may have to be modeled and the data interpreted in light of that model.''\footnote{\textcite[408]{boyd_evidence_2018}.} 
By following the process of data processing from beginning to end, we see how the EHT is particularly illustrative of the interdependence of data and models, or ``model-data symbiosis,'' as historian of science Paul Edwards has called it.\footnote{\textcite[281--282]{edwards_vast_2010}.} Rather than seeing models and data as separate, historians and philosophers of science have increasingly recognized the complicated relationships between the two. Not only are models data-laden, but data are also ``model-laden'' or ``model-filtered.'' Black hole imaging works with many different forms of model-laden data. Philosopher of science Alisa Bokulich has offered a helpful taxonomy of seven different ways in which data can be model-laden.\footnote{\textcite{bokulich_towards_2020}.} Although she illustrated each of her seven categories of model-ladenness with a different case from the geosciences, all categories are simultaneously represented in our case study of astronomical imaging in the EHT.

First, radio astronomy generally involves what Bokulich calls ``data conversion'': instead of capturing the data directly in the format of interest (an image of the sky brightness distribution), the instruments measure a proxy (interferometric patterns between pairs of antennas), which must then be converted into an image. VLBI poses a difficult problem of ``data interpolation'': filling in the gaps in very---in the EHT's case, extremely---sparse data. Correlation may be seen as a task of ``data fusion'': combining data from heterogeneous sources (distant telescopes) into a coherent and improved product (the correlated signals). Calibration consists of problems of ``data correction'': filtering many types of noise from the signals and correcting for various errors, not only from instruments but also from atmospheric effects. Reduction involves a sort of ``data scaling,'' since it compresses a wide frequency band into a single channel. And the Imaging Group used ``synthetic data'' for multiple purposes, including testing algorithms (during the Imaging Challenges) and selecting parameters (during ``fiducial'' imaging). Finally, the imaging process employed methods that constituted a kind of ``data assimilation'' by blending empirical and synthetic data into integrated results, namely images that look complete, clean, and sharp.

The combination of empirical and synthetic data into integrated results is not unique to black hole imaging. Philosophers of science have recently discussed cases of `data assimilation' in other fields. In climate science and meteorology, Wendy Parker has discussed atmospheric data assimilation in numerical weather prediction: computer simulations can fill spatial or temporal gaps in weather observations, such as missing measurements in particular places due to broken infrastructures, or estimates of weather for distant past periods (`retrospective analysis').\footnote{\textcite{parker_computer_2017}.} In particle physics, Margaret Morrison has discussed the role of simulations in the LHC experiments at CERN, for example in the discovery of the Higgs boson. She writes that ``simulation data are combined with signal data in the analysis of various events, rendering any sharp distinction between simulation and `experiment' practically meaningless.''\footnote{\textcite[289]{morrison_reconstructing_2015}.}

In both of these cases, the simulation models can be validated by comparing their outputs to empirical data, such as measurements from weather stations or particle detectors. Even if there are no available or trusted measurements for the very distant past or very high energies, the simulations can be checked with data from accessible time periods or energy regions. After the simulation models are validated on this basis, they may also generate outputs for the inaccessible periods or regions.\footnote{\textcite[18--26]{parker_computer_2017}; \textcite[305--312]{morrison_reconstructing_2015}.} Whether to trust such outputs is often a matter of dispute, but the simulations can be empirically tested at least for some situations. The EHT's situation differs: its methods could not be validated against empirical data at comparable resolution and data sparsity. Validation relied instead on synthetic data and on observations of other, less challenging targets such as the quasar 3C 279.
Although GRMHD simulations are well-grounded in trusted physical theories, namely general relativity and magnetohydrodynamics, their outputs could not be compared to any empirical data at these extreme limits of resolution and sparsity. Thus, the EHT advanced a distinctive form of data assimilation involving simulations whose validity in the relevant regime could not be independently tested.

This is an expression of a more general point.
The EHT pushed radio astronomy to its extreme limits of resolution, sometimes relying on innovative and unprecedented imaging methods involving various uses of simulation techniques and synthetic data. Since these methods could not be validated by comparing their outputs to any previously trusted observational images at the same resolution, EHT researchers resorted to alternative methods of validation. First, they tested their methods with synthetic data, including images generated by GRMHD simulations, which were theoretically well-grounded but not empirically tested. This is what happened at the Imaging Challenges (Section~\ref{blind}) and during ``fiducial imaging'' (Section~\ref{fiducial}). Moreover, they compared candidate images to \emph{each other} at different stages. This is what happened with ``blind comparison'' at the ``blind imaging'' stage (Section~\ref{blind}), and afterwards when images were shared across teams in the Imaging Group. While this did not amount to a properly independent verification, it was rather what one might call a working verification: comparing results across different teams was part of the working process of checking the images.\footnote{On the general challenge of building confidence in empirical results through iterative self-correction in the absence of independent benchmarks, see \textcite{chang_inventing_2004}.}

Finally, as we have also argued, the end product of the attempt at imaging black hole M87* (Figure~\ref{fig:m87-public}) could not have been attained without making \emph{aesthetic} choices. Indeed, the cropping of the image and the selection of the particular color map (Section~\ref{averaging}) were not conditioned by data, but rather by considerations such as the predicted associations that viewers, including particularly non-specialists, would have with different color maps and how this would influence their understanding of what they were looking at. Of course, such choices are hardly unique to the science of black hole imaging---yet, their role deserves to be highlighted, particularly as aspects of imaging are intrinsically artistic just as they are essential. This does remind us that the way that science shows us how the world is, and in particular what astronomy tells us the universe is like, is necessarily shaped by both data \emph{and} elements of artistry. Some EHT researchers have also sought to navigate the boundary between art and science, for instance when submitting their image to the photo collections of art museums, such as the MoMA in New York and the Rijksmuseum in Amsterdam, and celebrating its acceptance.\footnote{See the message by Heino Falcke on Twitter (now X) on January 15, 2020: \url{https://perma.cc/KB4T-PZY7}.}

The EHT produced important scientific evidence about black holes. But given the degree of uncertainty and flexibility involved at various stages of image-making, the full appreciation of the evidence requires seeing the range of variability among the multiple plausible images that result from different justifiable choices.\footnote{Recent philosophical discussions on the EHT have addressed the ``robustness'' of the final image. \textcite{galison_exploring_2023}, for example, has identified robustness as ``stability under adversarial challenge.'' On robustness in the EHT, see also \textcite{doboszewski_robustness_2024}; \textcite{weinstein_coincidence_2021}. On the broader research program in history, philosophy, and culture within the Next Generation Event Horizon Telescope (ngEHT), see \textcite{galison_next_2023}. On robustness in cosmology and high-energy physics, see \textcite{gueguen_robustness_2020}; \textcite{staley_securing_2020}. The EHT itself has derived epistemic confidence from the ``robustness'' of its result: ``A number of elements reinforce the robustness of our image and the conclusion that it is consistent with the shadow of a black hole as predicted by GR.'' \textcite[8]{event_horizon_telescope_collaboration_first_2019-1}.} As we have argued, the most valuable scientific evidence produced by the EHT comes not from the single image it ultimately advertised, but from the demonstration of the limited variability that emerged from the specific choices made. This leaves open the possibility that different choices, not tried or reported, could lead to greater variability. Yet the public communications of the EHT did not emphasize this variability. The imaging paper does discuss this variability to some extent, but mainly for the purpose of justifying some of the choices in the making of a definitive image. In fact, studying this variability demanded substantial effort on our part, and much of what we needed to know required going beyond the publications and looking into technical reports and source code.

Despite the multiplicity of plausible images, the EHT was committed to framing a single definitive image as its central result to a general public. This decision crucially shaped the process that we analyzed. Even when there were no clear justifications for a particular choice over reasonable alternatives, the EHT designed the process so that it would converge toward a single image. This was particularly evident at the last stage, when the results from three different methods were averaged into one image. The EHT used certain terms from the scientific vocabulary that may have served to generate trust in this result, such as ``ground truth,'' ``fiducial,'' and ``blind.'' Yet, as we have seen, in the process, it also inadvertently extended the meanings of these terms in unconventional ways. Using this terminology for scientific audiences may have played a similar rhetorical role to how the EHT used the terms ``photo'' and ``picture'' for popular audiences: they employed a choice of words with connotations of mechanical processes and objectivity, rather than focusing on elements of flexibility, choice, and artistry, even though the EHT also included a section on ``Image Uncertainties'' in its imaging paper.\footnote{\textcite[21--22]{event_horizon_telescope_collaboration_first_2019-4}.}

We have tried to nuance this depiction by presenting the epistemic and aesthetic choices in the process of image construction, and demonstrating the plurality of reasonable alternatives at each stage. We understand why the EHT ended up deemphasizing this plurality, as it might well have undermined its highly effective communication strategy. Yet, another way of presenting science is imaginable.
Researchers could present all reasonable alternatives instead of emphasizing convergence toward a single option. The central result need not be an individual image, but a full exploration of how different assumptions and choices affect the variability of plausible images. We believe this limited range of variability, and the careful process of data treatment and image production, to be more convincing than any single image.  
We recognize that such an approach might be a less effective way to communicate science to large public audiences as it would make the results seem less clear-cut and conclusive.\footnote{These tensions have also been observed in other disciplines, and between science communicators---journalists and scientists themselves---for at least a decade. The source of the oversimplification (or even exaggeration) of results to enlarge audiences and appear more impactful to journals and funding bodies has often been identified as the scientists themselves and their institutions; see, e.g., \textcite{sumner_association_2014}; \textcite{dejesus_generic_2019}.} Epistemologically, nonetheless, such a practice would improve and enrich the understanding of the evidence by scientists and wider publics alike.

\section*{Acknowledgments}

RO acknowledges support from the Dutch Research Council (NWO) through a Veni grant with file number VI.Veni.231S.057. This publication is part of the Dutch Black Hole Consortium with project number NWA.1292.19.202 of the research programme NWA, which is partly financed by NWO; JvD and ES acknowledge this support. The work presented here is also supported by the Carlsberg Foundation (grant CF23-1681), for which ES is grateful. The Niels Bohr Archive at the University of Copenhagen kindly hosts the animated and interactive figures. We thank EHT researchers who generously shared their recollections and insights, as well as audiences at CERN, the Utrecht Philosophy of Astronomy \& Cosmology seminar at Utrecht University, and the Black Hole Initiative at Harvard University. Finally, we are grateful to the journal's reviewers and editors, to colleagues in the History, Philosophy, \& Culture working group of the Next Generation Event Horizon Telescope (ngEHT), and to John D. Norton for their comments and suggestions.

\section*{Competing interests}

The authors have no competing interests to declare.

\printbibliography

\end{document}